\documentstyle[12pt,psfig]{article}
\input xy
\xyoption{all}

\begin{document}

\newcommand\M[2]{{\overline M}_{{#1},{#2}}}
\newcommand\th{\scriptstyle{{\rm th}}}

\hfill ILL-(TH)-04-5

\hfill hep-th/0406226

\vspace{0.5in}

\begin{center}

{\large\bf Notes on Certain (0,2) Correlation Functions }

\vspace{0.2in}

Sheldon Katz and Eric Sharpe \\
Departments of Mathematics and Physics, MC-382 \\
University of Illinois \\
Urbana, IL  61801 \\
{\tt katz@math.uiuc.edu}, {\tt ersharpe@uiuc.edu} \\

$\,$

\end{center}

In this paper we shall describe some correlation function
computations in perturbative heterotic strings that,
for example, in certain circumstances can lend themselves to
a heterotic generalization of
quantum cohomology calculations.  Ordinary quantum chiral rings
reflect worldsheet instanton corrections to correlation functions
involving products of elements of Dolbeault
cohomology groups on the target space.
The heterotic generalization described here involves computing
worldsheet instanton corrections to correlation functions defined
by products of elements of sheaf
cohomology groups.  One must not only compactify moduli spaces
of rational curves, but also extend a sheaf (determined by the gauge
bundle) over the compactification, and linear sigma models provide
natural mechanisms for doing both.  Euler classes of obstruction
bundles generalize to this language in an interesting way.

\begin{flushleft}
June 2004
\end{flushleft}

\newpage

\tableofcontents

\newpage

\section{Introduction}

In this paper we shall describe some correlation function computations 
in (large radius) heterotic nonlinear sigma models with K\"ahler 
(but not necessarily\footnote{As a result, these sigma models
have $(0,2)$ worldsheet supersymmetry, but beta functions do not
necessarily vanish.}
Calabi-Yau) targets, as part of a program to generalize
techniques developed to understand ordinary mirror symmetry,
to (0,2) mirrors.  We shall describe how to generalize the
rational-curve-counting of type II theories to
generalizations of the $\overline{ {\bf 27} }^3$ coupling
in heterotic strings.  We shall also check some of the
predictions made recently in \cite{abs}.

Recall that a perturbative heterotic string compactification
is defined by not only a manifold, but also a bundle over
that manifold.  We will only consider holomorphic vector
bundles over complex manifolds.
To fix notation, we shall denote the target space
by $X$, and we shall denote the holomorphic vector 
bundle\footnote{Occasionally one can use more general sheaves
than just locally-free sheaves \cite{dgm}; however, the circumstances
under which this is allowed are not well-understood at present,
so for simplicity throughout this paper we shall assume the gauge sheaf is
locally-free on $X$.}
on the target space $X$ by ${\cal E}$.
We shall assume throughout this paper 
that ${\cal E}$ possesses the following two properties:
\begin{enumerate}
\item $\Lambda^{top} {\cal E}^{\vee} \cong K_X$
\item $\mbox{ch}_2({\cal E}) = \mbox{ch}_2(TX)$
\end{enumerate}
The second of these properties is the well-known anomaly cancellation
condition in heterotic strings, and we shall see that it plays an
important role in defining the quantum product structure.
The first of these properties  
plays a different\footnote{
In \cite{bw} a condition analogous to this one appeared in making sense
of the path integral measure in certain theories used to examine
singlet correlation functions in heterotic theories.  There, however,
the analogue of ${\cal E}$ turned out to be the tangent bundle in
heterotic computations, and not the gauge bundle.
} and equally important role.
We shall see that it is required to make sense of product
structures (at both the classical and quantum levels),
as well as for the additive structure of the heterotic
chiral ring to be well-behaved.
Also, when the target space is Calabi-Yau, this condition guarantees
that a left-moving $U(1)$ symmetry (precisely the $U(1)_R$ on the $(2,2)$
locus) is nonanomalous.
The left-moving $U(1)$ in question is used to build representations of
the full low-energy gauge group.  
Hence for Calabi-Yau compactifications we preferentially consider
holomorphic vector bundles with vanishing first Chern class.
In any event, both of the properties above -- the anomaly cancellation
condition as well as the condition on first Chern classes -- will play an
important role in our analyses.

Technically, although we will sometimes speak of heterotic chiral `rings,'
following \cite{abs},
in general one will not always have a well-behaved
ring structure.  It would be more accurate to say that we will
study correlation functions between certain operators in
$(0,2)$ models, which in certain special cases ({\it e.g.} the gauge
bundle is a deformation
of the tangent bundle) can be interpreted
as corresponding to a ring structure.  If the gauge bundle is a deformation
of the tangent bundle, then this ring is a deformation of the
usual $(2,2)$ chiral ring.  The ring structure encodes the correlation 
functions in the same manner that the $(2,2)$ correlation functions are 
encoded in the quantum cohomology ring.

A related point concerns the possibility of loop corrections.
When the target is Calabi-Yau, then our three-point
correlation functions 
correspond to spacetime superpotential terms, and there is an old
nonrenormalization theorem based on Peccei-Quinn symmetries that
spacetime superpotentials do not receive loop corrections in
$\alpha'$, only worldsheet instanton corrections
(see {\it e.g.} \cite{dsww}).  
However, in this paper we are also interested in
massive two-dimensional theories in which the target space is not
Calabi-Yau, and we do not have a suitable generalization of the
nonrenormalization result to non-Calabi-Yau spaces.  Thus, in this paper
we compute worldsheet instanton corrections to certain correlation
functions, designed to generalize mathematical computations,
but sometimes, some of these correlation functions might also 
conceivably get loop corrections in addition.  As our purpose
in this note is to generalize some old mathematical computations,
we will not speak of this possibility further.

Another minor technical issue revolves around the spacetime
interpretation of our correlation functions.
When the target is Calabi-Yau, our three-point functions compute
superpotential terms.  However, the spacetime superpotential
is not a function, but rather a section of a line bundle over the
CFT moduli space \cite{edbagger}.  This fact is realized on the worldsheet
via $U(1)_R$ rotations as one moves about on CFT moduli space
\cite{distleritp}.  The `gauge choice' needed to uniquely define a 
superpotential term at any point in CFT moduli space therefore amounts
to a choice of normalization of the vertex operators, which cannot
be extended globally over CFT moduli space unless the line bundle
to which the superpotential couples is trivial.
We will not speak further of this issue in this paper.

We begin in section~\ref{halftwist} with a review of the half-twisted
(0,2) model in which we shall be computing correlation functions,
which reduces to the A model twist on the (2,2) locus.
In section~\ref{additive} we describe the states that will enter into
our correlation functions -- essentially, generalizations of
$\overline{ {\bf 27} }$'s.  In section~\ref{classical} we describe
classical (no worldsheet instantons) computations of the correlation
functions, and in section~\ref{quantum} we give a formal discussion
of how to take into account worldsheet instantons.
Our calculations generalize the rational curve counting of
type II theories, and in particular, the obstruction sheaf story
seems to generalize in an intriguing fashion.
To make sense of our formal calculations, we must compactify the
moduli space of worldsheet instantons and describe how to extend
certain sheaves (induced by the gauge bundle) over the compactification.
In section~\ref{lsmmod} we review how linear sigma models can be used
to provide a natural compactification, and furthermore how they
naturally define and extend the relevant sheaves, in a fashion compatible
with symmetries.  Finally in section~\ref{checkabs} we apply this
technology to check some predictions of \cite{abs}.

We should mention some related work.
For example, considerations of $(2,2)$ sigma models on supermanifolds
\cite{schwarz} lead one to study holomorphic vector bundles on
Calabi-Yau manifolds, and their extensions over moduli spaces of 
rational curves.
Some related work in heterotic compactifications is
\cite{bw,evaed,candelasetal,basusethi}, where correlation functions
of gauge singlets were studied (and shown to vanish).
In this paper, we are concerned with heterotic correlation functions
between vertex operators for charged states -- no gauge singlets appear
in our correlation functions, and so the vanishing results of
\cite{bw,evaed,candelasetal,basusethi} are not relevant.

Finally, it should be noted throughout this paper that we will
{\it not} couple our models to worldsheet gravity -- our computations
are performed for a fixed complex structure on the worldsheet --
and furthermore, we will only consider genus zero worldsheets.

\section{Half-twisted $(0,2)$ theory}   \label{halftwist}

In typical quantum cohomology calculations,
the worldsheet theory is the A-model topological field theory.
A $(0,2)$ theory cannot be twisted to give a completely topological
theory, but we can do most of the twisting, and will recover most of
the corresponding results.

In a standard $(0,2)$ theory, one has right-moving fermions $\psi_+$ coupling
to $\sqrt{K_{\Sigma}} \otimes \phi^* TX$ and left-moving
fermions $\lambda_-$ coupling to $\sqrt{\overline{K}_{\Sigma}} \otimes
\phi^* \overline{{\cal E}}$, where $\Sigma$ denotes the worldsheet.  
We define the ``half-twisted theory'' by coupling fermions to bundles
as follows:\footnote{We have used Hermitian metrics on $X$ and ${\cal E}$
to write $\phi^*T^{0,1}(X)$ as $(\phi^*T^{1,0}(X))^\vee$ and
$\phi^*{\cal E}$ as $(\phi^*\overline{{\cal E}})^\vee$.  We do this to
make a Riemann-Roch calculation in Section~\ref{classical} more transparent.}
\begin{eqnarray*}
\psi_+^i & \in & \Gamma_{ C^{\infty} }\left( \phi^* T^{1,0}X \right) \\
\psi_+^{ \overline{\imath}} & \in & \Gamma_{ C^{\infty} }\left( 
K_{\Sigma} \otimes \left( \phi^*T^{1,0}X \right)^{\vee} \right) \\
\lambda_-^a & \in & \Gamma_{ C^{\infty} }\left( 
\overline{K}_{\Sigma} \otimes \left( \phi^* \overline{{\cal E}} \right)^{\vee}
 \right) \\
\lambda_-^{ \overline{a} } & \in & \Gamma_{ C^{\infty} }\left(
\phi^* \overline{ {\cal E} } \right)
\end{eqnarray*}
and it is this half-twisted theory that we shall be studying throughout
this paper.  
In other words, just as in a standard topological twist,
we make the worldsheet spinors into worldsheet scalars and vectors.
Note that although we perform this operation on both left- and right-movers,
the resulting theory will not be a topological field theory in general,
since it only has $(0,2)$ worldsheet supersymmetry, and not $(2,2)$.
Also note that on the $(2,2)$ locus, where ${\cal E} = TX$,
the twisting above is equivalent to the A model twist.
See \cite{evaed} for another discussion of half-twisted theories in
heterotic compactifications.

One of the reasons two-dimensional topological field theories are useful
is that they simplify the computation of correlation functions
(see {\it e.g.} \cite{antoniadisetal}).
Many three-point functions in physical (type II) theories are the same
as certain three-point functions in corresponding topological field
theories.  Schematically,
\begin{displaymath}
< \psi \psi \phi >_{phys} \: = \: < \psi \psi \psi >_{tft}
\end{displaymath}
In a nutshell, 
the reason for this equivalence is that the spectral flow insertion
used to generate the vertex operator for the spacetime boson $\phi$ above
from the vertex operator for the corresponding spacetime spinor 
can also be interpreted as generating the topological field theory.

Although the mechanism for relating physical and topological correlation
functions is well-known, let us pause for a moment to very briefly 
and schematically review it.
In a nutshell, to twist the worldsheet theory means adding
a term proportional to $\int_{\Sigma} \frac{1}{2} \omega \overline{\psi}
\psi$ to the worldsheet action, where $\omega$ is the worldsheet
spin connection.  Now, $\overline{\psi}\psi$ is proportional to the
$U(1)_R$ current $J$.  If we bosonize $J \sim \partial \phi$,
then the term added looks like $\int_{\Sigma} R \phi$.
By concentrating the worldsheet curvature at points,
so that $R \sim \delta^2(z-z_0)$, we see that topological twisting
is essentially the same as inserting factors of the form
$\exp(\phi)$,
which is spectral flow.

Analogous statements can also be made relating correlation functions
in physical $(0,2)$ theories to corresponding correlation functions
in the half-twisted theory described above.
Recall that a three-point correlation function in a physical
$(0,2)$ theory has two aspects, coming from the left- and right-movers.
In a three-point function the right-movers behave just as in a type II theory,
as described above.
The left-movers encode gauge
information, and require an analogue of spectral flow, which can also be
alternately interpreted as twisting.  
For example, suppose we have an irreducible rank three 
holomorphic vector bundle, and we wish to compute a 
$\overline{{\bf 27}}^3$ coupling\footnote{On the $(2,2)$ locus,
the $\overline{{\bf 27}}^3$
gets worldsheet instanton corrections, whereas the ${\bf 27}^3$ does
not.  On the $(2,2)$ locus, the $\overline{{\bf 27}}^3$ corresponds to
an A model calculation, whereas the ${\bf 27}^3$ corresponds to a B model
computation.}.  The bundle breaks $E_8$ to $E_6$, and the
$E_6$ is built on the worldsheet from $SO(10) \times U(1)$,
the $SO(10)$ from the left-moving fermions in the $E_8$ but not
coupled to the bundle, and the $U(1)$ from a left-moving
symmetry\footnote{It is straightforward to compute that in a physical
$(0,2)$ theory on a Calabi-Yau, this left-moving $U(1)$ will be
nonanomalous precisely when $c_1({\cal E}) = 0$.}
on the fermions coupled to the bundle,
under which $\lambda^a$ and $\lambda^{\overline{a}}$ have
equal and opposite charges.  (On the $(2,2)$ locus,
this $U(1)$ is the $U(1)_R$ of the left-moving ${\cal N}=2$ algebra.)
Just as a physical three-point correlation function involves both spacetime
spinors and a spacetime boson, one analogously uses both the
$\overline{{\bf 16}}$ and ${\bf 10}$ 
representations of $SO(10)$ that are part of
the $\overline{{\bf 27}}$ of $E_6$: 
\begin{displaymath}
\overline{{\bf 27}} \: = \: {\bf 10}_{-1} \oplus \overline{{\bf 16}}_{1/2}
\oplus {\bf 1}_2
\end{displaymath}
under $SO(10) \times U(1)$.
In particular, the physical $\overline{{\bf 27}}^3$ correlation function
can be computed in the form
\begin{displaymath}
< \psi_{ \overline{{\bf 16}} } \psi_{ \overline{{\bf 16}} } \phi_{ {\bf 10} } >_{phys}.
\end{displaymath}
Just as in the right-movers, the vertex operator for $\phi$ could be
obtained from that for $\psi$ via spectral flow, similarly the
left-moving ${\bf 10}$ can be obtained from the left-moving ${\bf 16}$
through the action of a left-moving analogue.
Just as in $(2,2)$ theories, where these spectral flow insertions could
be interpreted as giving topological twists, in $(0,2)$ theories
these insertions of (analogues of) spectral flow can also be
interpreted as giving the half-twisted theory, hence
\begin{displaymath}
< \psi_{ \overline{{\bf 16}} } \psi_{ \overline{{\bf 16}} } \phi_{ {\bf 10} } >_{phys}
\: = \: < \psi \psi \psi >_{half-twist}.
\end{displaymath}

Because the half-twisted theory simplifies computations of correlation
functions as described above, in this paper we shall use it
exclusively to compute correlation functions.

In computations of rational
curve corrections in type II theories, the
A model is often coupled to worldsheet topological gravity in the literature.  
In this paper,
we will not couple to topological gravity.  Furthermore, when computing
$n$-point correlation functions on ${\bf P}^1$ for $n>3$, we shall not
use descendants of $n-3$ of the vertex operators; we shall compute products
in the purely half-twisted $(0,2)$ theory.

\section{Relevant massless states in (0,2) theories}   \label{additive}

In a type II string theory, the chiral ring consists of massless
RR sector states, which (at large radius) are counted by Dolbeault cohomology
$H^{p,q}(X)$ of the target space $X$, or, equivalently,
elements of the sheaf cohomology groups
\begin{displaymath}
H^q\left(X, \Lambda^p T^* X \right).
\end{displaymath} 
By Hodge theory, these Dolbeault cohomology groups generate the full
de Rham cohomology.

In a heterotic string compactification on a space $X$ with 
gauge bundle ${\cal E}$,
recall from \cite[section 3]{dg} that the charged massless RR states  
are counted (at large radius) by
\begin{displaymath}
H^q\left(X, \Lambda^p {\cal E}^{\vee} \right)
\end{displaymath}
which is clearly analogous to the type II string result.
In the special case that ${\cal E} = TX$, we duplicate the
type II result.

We shall sometimes, when sensible, speak of a heterotic chiral ring
associated 
to certain $(0,2)$
heterotic theories.  This ring will be described additively by the sum
of sheaf cohomology groups of the form above, {\it i.e.}
\begin{displaymath}
H^{*,*}_{het} \: \equiv \: \sum_{p,q} 
H^q\left(X, \Lambda^p {\cal E}^{\vee} \right)
\end{displaymath}  
Note that this ring is naturally bigraded.  Physically, that bigrading
corresponds to the distinction between left- and right-movers.  The heterotic
chiral ring is associated to part of the gauge sector of the heterotic
theory.  The multiplication in the ring is described by the
correlation functions.

In principle, there are additional massless states in a large radius
heterotic compactification -- there are gauge singlets that correspond
to complex, K\"ahler, and bundle moduli, as well as additional charged
matter fields, such as elements of 
$H^n\left( X, \mbox{End } {\cal E}\right)$
come from the NS-R sector.  However, we shall not consider
such fields here.

Clearly we have states of the form $H^1(X, {\cal E}^{\vee})$ in our
state space, 
the $(0,2)$ generalization of $H^1(T^{\vee})$, corresponding to K\"ahler
moduli on the $(2,2)$ locus.  
Note that the $(0,2)$ theory does still contain K\"ahler
moduli, but in the $(0,2)$ theory those states are strictly gauge
neutral.  A $(2,2)$ theory contains both gauge neutral as well as
gauge charged states corresponding to complex structure moduli -- {\it
i.e.} in addition to singlet moduli fields in the target space theory
corresponding to complex moduli,
there are also charged ${\bf 27}$'s corresponding to complex moduli.
A $(0,2)$ theory contains singlets corresponding to complex moduli,
but even for a rank three ${\cal E}$, the ${\bf 27}$'s are counted by
$H^1(X, {\cal E})$ and not $H^1(X, T)$.

In the introduction we mentioned that, in addition to the
anomaly cancellation condition, throughout this paper we shall
impose the constraint that
\begin{equation}
\label{u1cond}
\Lambda^{top} {\cal E}^{\vee} \: \cong \: K_X.
\end{equation}
We can see one motivation for this constraint already in the
additive structure of our chiral states.
Recall that Serre duality in $(0,2)$ theories maps spectra back
into themselves \cite{dg}.
Serre duality acts as \begin{eqnarray*}
H^i\left(X, \Lambda^j {\cal E}^{\vee} \right) & \cong &
H^{n-i}\left(X, \Lambda^j {\cal E} \otimes K_X \right)^* \\
& \cong & H^{n-i}\left(X, \Lambda^{r-j} {\cal E}^{\vee} \otimes
\Lambda^r {\cal E} \otimes K_X \right)^*
\end{eqnarray*}
where $r$ is the rank of ${\cal E}$.
Notice that when the line bundle
\begin{displaymath}
\Lambda^r {\cal E} \otimes K_X
\end{displaymath}
on $X$ is trivial, {\it i.e.} $\Lambda^r {\cal E}^{\vee} \cong K_X$,
Serre duality still closes our states back into
themselves, but if this line bundle is nontrivial, Serre duality does
not close the states back into themselves.
Thus, just to make the additive structure well-behaved under
Serre duality, we must require the constraint above.

The condition (\ref{u1cond}) also impacts correlation functions.
For definiteness, let $E$ have rank $r$ and let $X$ have dimension $n$.
Note that the condition 
$\Lambda^{r} {\cal E}^{\vee} \cong K_X$ implies that there is an 
isomorphism
\[
\phi:H^n(X,\Lambda^{r} {\cal E}^{\vee})\to {\bf C}.
\]
This isomorphism already
allows classical
products of operators in the chiral ring to be evaluated, and we see
that it will similarly be crucial in the evaluation of the quantum
products.  

For Calabi-Yau compactifications, another physical motivation for this
constraint comes from the fact that our bundles must be embeddable
inside $E_8$.  Although $SU(n)$ has natural $E_8$ embeddings for
small $n$ which cleanly correspond to worldsheet physics,
realizing $U(n)$ embeddings on the worldsheet 
is not as well understood. In particular, for a $U(n)$ embedding
there is  
a worldsheet anomaly 
in a left-moving $U(1)$ symmetry used to build vertex operators.
Thus, although one can work with $(0,2)$ CFT's with $c_1 \neq 0$,
to make them useful for a compactification one typically adds extra left-movers
to cancel out the $c_1$, hence as a practical matter we only consider
holomorphic vector bundles with $c_1=0$.

\section{Classical correlation functions}   \label{classical}

In a $(2,2)$ chiral ring, the classical product is obtained
by wedging together enough differential forms to get a top form
on the target space $X$, which can then be
integrated over $X$ to produce a number.
In other words, for the product of $k$ operators we have the maps
\begin{equation}  \label{classprod}
H^{q_1}(X, \Lambda^{p_1} T^* X) \otimes \cdots \otimes
H^{q_k}(X, \Lambda^{p_k} T^* X) \: \longrightarrow \:
H^n(X, \Lambda^n T^*X) \: \cong \: {\bf C}
\end{equation}
where $\sum q_i = \sum p_i = n$. The first map is given by
cup and wedge products, and the second map is given by integration
of a top form on $X$.
This mathematical analysis is backed-up physically
by {\it e.g.} ghost number conservation, which implies that the
only nonzero correlation functions can come from top-degree forms.

The classical correlation functions in our heterotic theory involve
a product of states such that the sums of the degrees
of the sheaf cohomology groups is the dimension $n$ of the target-space
$X$, and that the number of ${\cal E}$'s in the coefficients
equals the rank $r$ of ${\cal E}$.  
Indeed, wedging together representative differential forms,
from such a product we get an element of 
\begin{equation}
\label{topclass}
H^n\left(X, \Lambda^r {\cal E}^{\vee} \right).
\end{equation}
This is the analogue of a top-degree form in the present case.

However, we can only get a number from this top-degree form in
special circumstances.  In the special case that
$\Lambda^r {\cal E}^{\vee} \cong K_X$, then we can 
identify an element of (\ref{topclass}) with a top form on $X$, which
can then be integrated to get a number.

Next, let us take a moment to consider the physical situation
more carefully. 

In the half-twisted theory, as described in section~\ref{halftwist},
the $\psi_+^i$ couple to $\phi^* T^{1,0}X$
and the $\psi_+^{ \overline{\imath}}$ couple to 
$K_{\Sigma} \otimes \left( \phi^* T^{1,0}X\right)^{\vee}$, 
hence the number of $\psi_+^i$
zero modes, 
minus the number
of $\psi_+^{\overline{\imath}}$ zero modes, is given by
Hirzebruch-Riemann-Roch as
\begin{displaymath}
\chi\left( \phi^* T^{1,0}X \right)  \: = \:
c_1( \phi^* TX) \: + \: n(1-g)
\end{displaymath}
This is the anomaly in the (right-moving) $U(1)_R$ symmetry.
(Compare {\it e.g.} \cite{abs} above equation~(154).)

This anomaly gives us a selection rule that is responsible
for the top-degree-form constraint when there are no worldsheet instantons.
Consider the case that our vacuum consists of constant maps $\phi$,
{\it i.e.} no worldsheet instantons.  Then $\phi^* TX$ is a trivial
bundle over the worldsheet, so $c_1(\phi^* TX) = 0$,
hence $\chi = n (1-g)$.  Thus at string tree level there are $n$
$\psi_+^i$ zero modes, hence the degrees of
the sheaf cohomology groups must add up to $n$ ( $= \mbox{dim }X$)
in order to hope to get a nonzero result for the correlation function.

As outlined in section~\ref{halftwist}, 
to most efficiently see the product structure
in the chiral ring, we also want to consider the case that the
left-moving fermions are also twisted, so that $\lambda_-^a$ couples
to $\overline{K}_{\Sigma} \otimes \left( \phi^* \overline{{\cal E}} 
\right)^{\vee}$ and 
$\lambda_-^{ \overline{a}}$ couples to
$\phi^* \overline{ {\cal E} }$.
Here we compute an anomaly
\begin{displaymath}
\chi\left( \phi^* {\cal E} \right) \: = \:
c_1\left( \phi^* {\cal E} \right) \: + \:
r(1-g).
\end{displaymath}
When we are expanding about constant $\phi$, we have
$c_1(\phi^* {\cal E}) = 0$.  Thus at string tree level we recover
the rule that for a correlation function to be nonvanishing,
the number of $\lambda_-^{\overline{a}}$'s 
appearing must equal the rank of ${\cal E}$.

In any event, from counting zero modes of worldsheet fermions, 
we see that when there are no
worldsheet instantons, for a product to be nonvanishing the sum
of the degrees of the sheaf cohomology groups must match the
dimension of the target, and the sum of the exterior powers of the
coefficients must match the rank of ${\cal E}$.  In other words,
for a product to be non-vanishing, the wedge product of differential
forms must give us the analog of a top form, as we outlined earlier.

As noted above, 
in the special case that $\Lambda^r {\cal E}^{\vee} \cong K_X$,
this wedge product living in (\ref{topclass}) can be identified with a
top form and subsequently integrated to get a number.
A precise choice of trivialization of $\Lambda^r {\cal E}
\otimes K_X^{\vee}$ is needed to normalize correlation functions.
This is analogous to the $(2,2)$ chiral ring, where a precise choice of
nowhere-zero holomorphic top form is needed to normalize the correlation
functions.

Thus, the constraint $\Lambda^r {\cal E}^{\vee} \cong
K_X$ not only makes the path integral well-defined,
but is also used in order to be able to define the classical product.
We will see the same constraint in defining quantum products.

\section{Quantum correlation functions -- formal discussion}
\label{quantum}

\subsection{Generalities}

The selection rules derived above from anomalies no longer apply in
the presence of worldsheet instantons,
but rather are corrected.  Previously if the sum of the degrees of
the sheaf cohomology groups was greater than the dimension of the
manifold, the resulting correlation function must vanish,
reflecting a basic property of the (`classical') Yoneda product,
really just a property of wedge products. 
In an instanton background, however, it is possible to have
a nonzero correlation function even when that sum of degrees is
strictly greater than the dimension of the target manifold.
This is what gives rise to ``quantum'' products -- products that
are nonzero, but become zero when $\alpha' \rightarrow 0$.

Suppose that we want to compute a correlation function associated to
elements of the heterotic chiral ring.
Specifically, recall that the selection rule on right-movers is that
the sum of the degrees of the sheaf cohomology groups must equal
\begin{equation}   \label{shfdeg}
c_1(\phi^* TX) + n (1-g)
\end{equation}
where $n$ is the dimension of the target and $g$ is the genus of the
worldsheet,
and the sum of the powers of the bundle ${\cal E^\vee}$ appearing must be
\begin{equation}    \label{bdle}
c_1(\phi^* {\cal E}) \: + \: r (1-g)
\end{equation}
where $r$ is the rank of ${\cal E}$.
In a one-instanton background, neither $c_1(\phi^* TX)$ nor
$c_1(\phi^* {\cal E})$ need vanish.

There is a significant complication,
not present in the $(2,2)$ case, coming from operator determinants.
In a $(2,2)$ topologically-twisted theory,
the operator determinants for the left- and right-moving worldsheet
fermions precisely cancel against the operator determinant for the
worldsheet bosons.  In a physical theory, no such cancellation occurs,
and certainly in a $(0,2)$ theory, twisted or not,
no such cancellation is possible.  (See also \cite{candelasetal} for a more
extensive discussion of this issue.)
Thus, all string tree level correlation functions necessarily
contain numerical factors amounting to the partition function
of left-moving fermions on $S^2$.

If these numerical factors were the same for all worldsheet instanton
contributions, then we could ignore them, or absorb them into
redefinitions.
However, these numerical factors for any one worldsheet instanton
depend upon the restriction of ${\cal E}$ to that rational curve,
as after all they represent the partition function for left-moving
fermions coupling to ${\cal E}$.  If two different rational curves
have non-isomorphic restrictions of ${\cal E}$, then 
the numerical factors will be different.

Far worse now is the case of a family of curves,
in which the cohomology of the restriction of ${\cal E}$ can jump 
in the family \cite{okoneketal}.
If the cohomology of the restriction of ${\cal E}$ jumps, then 
surely the corresponding
numerical factor jumps, and so the required integral over the
moduli space of zero modes can not have nearly so simple a form
as appeared in \cite{pauldave}.

We shall ignore this possibility, and work only in $g=0$ (at fixed complex
structure), so that this determinant is just a number that can be
reabsorbed into other definitions.  
We will also ignore group theory
factors associated to the representations of the unbroken gauge group
needed to specify the gauge sector contributions to the operators.

In the rest of this section we shall only work formally, on
smooth not necessarily compact spaces of maps of fixed degree from the 
worldsheet into $X$ of fixed degree.
Our goal in this section is to outline the general ideas,
not to fill in all details.
In section~\ref{lsmmod} we shall describe how to compactify the
moduli spaces and extend these constructions over those compactifications.

\subsection{Integration over moduli space}   \label{quant:int}

In this subsection, we shall formally outline how to compute
rational curve corrections in the special case that there
are no excess zero modes, that there is no analogue of the
obstruction sheaf.  (We shall make this restriction precise
momentarily.)  The general case will be discussed in 
section~\ref{excesszeromodes}.
Readers unfamiliar with these computations for $(2,2)$ theories
might wish to consult \cite[section 3.3]{ed1}, \cite[section 3.3]{ed2}
for a very readable review.

In this section we shall only work formally.  We consider an idealized
situation which is almost never realized:  we suppose that we have
a smooth and compact instanton moduli space ${\cal M}$ of maps from the fixed
worldsheet $\Sigma$ into $X$ of fixed degree.  On the other hand, it is quite
common to find noncompact but smooth moduli spaces.  In section~\ref{lsmmod}
we shall describe how to compactify the moduli spaces and extend these
constructions over those compactifications.

Recall in the $(2,2)$ case, the chiral ring is composed of $(p,q)$ forms
on the target space $X$.  There is a selection rule that correlation functions
are nonzero when the sum of the $p$'s and $q$'s separately equals
\begin{displaymath}
\left( \mbox{dim } X \right) (1-g) \: + \: c_1\left( \phi^* TX \right)
\end{displaymath}
(the same as the selection rule~(\ref{shfdeg}) on the degrees of
sheaf cohomology groups).
Each of those $(p,q)$ forms on $X$ is associated to a $(p,q)$ form
on the moduli space ${\cal M}$ of worldsheet instantons of given degree,
and the selection rule above becomes the statement that the
wedge product of those differential forms on ${\cal M}$ is a formally
a top form,
since formally the dimension of ${\cal M}$ is given by
\begin{displaymath}
\left( \mbox{dim } X \right) (1-g) \: + \: c_1\left( \phi^* TX \right).
\end{displaymath}

In the heterotic case, with a rank $r$ bundle ${\cal E}$ obeying $\Lambda^r
{\cal E}^{\vee} \cong K_X$ as well as the anomaly cancellation condition,
elements of the sheaf cohomology groups
\begin{displaymath}
H^p\left(X, \Lambda^q {\cal E}^{\vee} \right)
\end{displaymath}
are mapped to sheaf cohomology groups
\begin{displaymath}
H^p\left({\cal M}, \Lambda^q {\cal F}^{\vee} \right)
\end{displaymath}
on the moduli space ${\cal M}$, where ${\cal F}$ is a sheaf on ${\cal M}$
of rank $\Gamma(X, \phi^* {\cal E})$.  We will discuss the precise
map in the next subsection, but for the purpose of
outlining the general ideas, for now we merely assert such a map exists.

Let us assume that we have a 
universal instanton $\alpha: \Sigma \times {\cal M} \rightarrow X$,
where $\Sigma$ is the worldsheet.\footnote{Here is where we will find it hard
to enforce the compactness of ${\cal M}$ in practice.
In a typical situation maps from the worldsheet can degenerate to maps
from the worldsheet with some 2-spheres attached.}  
We have the natural projection 
$\pi: \Sigma \times
{\cal M} \rightarrow {\cal M}$.  Now we can define a sheaf ${\cal F}$ 
on ${\cal M}$ by
\begin{displaymath}
{\cal F} \: = \: \pi_* \alpha^* {\cal E}.
\end{displaymath}
In this section, we will assume that
\begin{equation}  \label{noexcess}
R^1 \pi_* \alpha^* {\cal E} \: = \: 0 \: = \: R^1 \pi_* \alpha^* TX.
\end{equation}
In the case where ${\cal E} = TX$, we have that ${\cal F} = T {\cal M}$,
as the fiber of ${\cal F}$ at a point of ${\cal M}$ corresponding to an
instanton $\phi: \Sigma \rightarrow X$ is $H^0(\Sigma, \phi^* TX)$,
the space of first-order deformations of the map $\phi$.\footnote{Note that
we are {\it not} considering stable maps here as we are not identifying
maps differing by an automorphism of $\Sigma$.  
As we have not coupled to worldsheet gravity, worldsheet automorphisms
are irrelevant.  Unlike topological {\it string} theories,
in a topological field theory, even at genus zero one can have
nonzero two-point couplings. }
The first equality in~(\ref{noexcess}) tells us that the map
$\phi$ is unobstructed.  In particular ${\cal M}$ is smooth.  The second
equality is the natural extension.  By Riemann-Roch, this condition implies
that the fibers of ${\cal F}$ have the same dimension,
so that ${\cal F}$ is a vector bundle of rank given in equation~(\ref{bdle}).

Physically the assumption~(\ref{noexcess}) means that there are no
excess zero modes.  We will discuss the more general case in 
section~\ref{excesszeromodes}. 

Under the assumption~(\ref{noexcess}), we can compute $c_1({\cal F})$ by
Grothendieck-Riemann-Roch.  Letting a subscript of $k$ on a cohomology
class denote its complex codimension $k$ component, and letting $\eta$
be the pullback to $\Sigma\times {\cal M}$ of the cohomology class of a
point of $\Sigma$, we have
\begin{equation}  \label{grr1}
c_1({\cal F}) \: = \: \mbox{ch}({\cal F})_1 \: = \:
\pi_*\left( \left( \mbox{ch}(\alpha^* {\cal E}) \mbox{Td}(T \Sigma) \right)_2
\right) \: = \: \pi_*\left( \alpha^* \left(c_1({\cal E})^2/2-
c_2({\cal E}) \right)+\left(1-g\right)\eta\alpha^*c_1({\cal E})\right).
\end{equation}
In~(\ref{grr1}) we have used 
$\mbox{ch}_2({\cal E}) = c_1^2({\cal E})/2 - c_2({\cal E})$.

If we now apply~(\ref{grr1}) to ${\cal E} = TX$ we obtain
\begin{displaymath}
c_1(T {\cal M}) \: = \: 
\pi_*\left( \left( \mbox{ch}(\alpha^* TX) \mbox{Td}(T \Sigma) \right)_2
\right) \: = \: \pi_*\left( \alpha^* \left(c_1(TX)^2/2-
c_2(TX) \right)+\left(1-g\right)\eta\alpha^*c_1(TX)\right).
\end{displaymath}
By the anomaly condition together with (\ref{u1cond}), we conclude that 
$c_1({\cal F}) = c_1(T {\cal M})$,
implying\footnote{At least when ${\cal M}$ is smooth, K\"ahler,
and simply-connected,
assumptions we will freely make for the purposes of the formal discussion
of this section.} that $\Lambda^{top} {\cal F}^{\vee} \cong K_{ {\cal M} }$,
as desired.

The selection rules~(\ref{shfdeg}) and (\ref{bdle}) then imply that
we can get nonvanishing correlation functions by a straightforward extension
of the procedure~(\ref{classprod}), replacing $X$ by ${\cal M}$
and $T^*X$ by ${\cal F}^{\vee}$:
\begin{equation}   \label{quanthetprod}
H^{q_1}( {\cal M}, \Lambda^{p_1} {\cal F}^{\vee} ) \otimes \cdots
\otimes H^{q_k}({\cal M}, \Lambda^{p_k} {\cal F}^{\vee}) \: \longrightarrow
\: H^N\left({\cal M}, \Lambda^R {\cal F}^{\vee}\right) \: \cong \:
H^N\left( {\cal M}, K_{ {\cal M} } \right) \: \cong \: {\bf C},
\end{equation}
given by cup and wedge product of cohomology classes followed by
integration of a top form.

Technically, as mentioned in the previous section,
the correlation function should be more properly described as
\begin{displaymath}
\int_{ {\cal M} } \left( \frac{ \mbox{det } \overline{\partial}_{ \phi^* {\cal E} } }{
\mbox{det } \overline{\partial}_{ \phi^* TX } } \right) 
H^{top}\left( {\cal M}, K_{ {\cal M} } \right)
\end{displaymath}
instead of merely
\begin{displaymath}
\int_{ {\cal M} } H^{top} \left( {\cal M}, K_{ {\cal M} } \right).
\end{displaymath}
But at genus zero, the ratio of operator determinants is just a number,
which we shall suppress.  We shall also studiously ignore the possibility
that the splitting behavior of $\phi^* {\cal E}$ changes over ${\cal M}$
in such a way that this ratio changes on a set of nonzero measure
on ${\cal M}$.  Understanding this ratio of operator determinants
would be much more important for calculations at higher genus, and for
properly understanding coupling to worldsheet gravity.

In~(\ref{quanthetprod}), $N$ is the dimension of ${\cal M}$ as given by
the selection rule~(\ref{shfdeg}) and $R$ is the rank of ${\cal F}$
as given by the selection rule~(\ref{bdle}).
Also, note that the anomaly cancellation condition played a crucial
role in defining this product, as it is only because of the anomaly
cancellation condition that we have $\Lambda^R {\cal F}^{\vee}
\cong K_{ {\cal M} }$.

In the next section, we will describe maps
\begin{displaymath}
\psi_k: \: H^q\left(X, \Lambda^p {\cal E}^{\vee} \right)
\: \longrightarrow \:
H^q\left( {\cal M}, \Lambda^p {\cal F}^{\vee} \right),
\end{displaymath}
the subscript $k$ corresponding to the $k^{\th}$ insertion point.  
This will complete our description of the correlation functions in our
ideal situation.  Given cohomology classes $\eta_i \in
H^{q_i}\left(X, \Lambda^{p_i} {\cal E}^{\vee} \right)$ satisfying
the selection rules $\sum q_i = N$ and $\sum p_i = R$, we 
apply~(\ref{quanthetprod}) to the $\psi_i ( \eta_i )$ to get the
value of the desired correlation function, using algebraic geometry.

\subsection{Evaluation map}

In order to make sense of the calculations just described,
we need to describe the maps $\psi_i: H^q(X, \Lambda^p {\cal E}^{\vee})
\rightarrow H^q({\cal M}, \Lambda^p {\cal F}^{\vee})$.

More precisely, by restricting the universal map $\alpha: \Sigma
\times {\cal M} \rightarrow X$ to $p_i \in \Sigma$
(where $p_i$ is the $i^{\th}$ insertion point),
we have an evaluation map
\begin{displaymath}
\mbox{ev}_i: \: {\cal M} \: \longrightarrow \: X, \: \: \:
\mbox{ev}_i(\phi) \: = \: \phi(p_i).
\end{displaymath}
Note that upon identifying ${\cal M}$ with $\{ p_i \} \times
{\cal M}$, we see that $\mbox{ev}_i$ is just the restriction of
$\alpha$ to ${\cal M}$.
We need to define a map
\begin{equation}    \label{eval1}
\psi_i: \: H^q\left( X, \Lambda^p {\cal E}^{\vee} \right) \: 
\mapsto \: H^q\left( {\cal M}, \Lambda^p {\cal F}^{\vee} \right).
\end{equation}

First, note that pulling back by $\mbox{ev}_i$, we get a map
\begin{equation}   \label{pull1}
H^q\left(X, \Lambda^p {\cal E}^{\vee} \right) \: \longrightarrow \:
H^q\left( {\cal M}, \Lambda^p \left( \alpha^* {\cal E} \right)^{\vee} |_{
\{ p_i \} \times {\cal M} } \right).
\end{equation}
So to derive the map~(\ref{eval1}) we just need to find
a sheaf map
\begin{displaymath}
\Lambda^p \left( \alpha^* {\cal E} \right)^{\vee} |_{ \{ p_i \} \times {\cal M} } 
\: \longrightarrow \:
\Lambda^p {\cal F}^{\vee}
\end{displaymath}
yielding a corresponding map on cohomology
\begin{displaymath}
H^q\left( {\cal M}, \Lambda^p \left( \alpha^* {\cal E} \right)^{\vee}
|_{ \{ p_i \} \times {\cal M} } \right) \: \longrightarrow \:
H^q\left( {\cal M}, \Lambda^p {\cal F}^{\vee} \right)
\end{displaymath}
to compose with the pullback map~(\ref{pull1}) to arrive at $\psi_i$.

For simplicity of notation, we complete the argument in the case
$p=1$; the generalization to arbitrary $p$ is straightforward.
We shall demonstrate the existence of the dual map ${\cal F} \rightarrow
\left( \alpha^* {\cal E} \right) |_{ \{ p_i \} \times {\cal M} }$, 
and will work in terms of
local sections of the corresponding sheaves.  Let $U$ be an open
subset of ${\cal M}$, then
\begin{eqnarray*}
{\cal F}(U) & = & \left( \pi_{*} \alpha^* {\cal E} \right) (U)
\mbox{ (by definition)} \\
& = & \left( \alpha^* {\cal E} \right) ( \pi^* U ) \\
& = & H^0 \left( \Sigma \times U, \alpha^* {\cal E} \right)
\end{eqnarray*}
so we have now expressed local sections of ${\cal F}$ in terms of
local sections of $\alpha^* {\cal E}$.  Next, we can restrict those
sections, getting a map
\begin{displaymath}
H^0\left( \Sigma \times U, \alpha^* {\cal E} \right) \: \longrightarrow \:
H^0 \left( U, \left( \alpha^* {\cal E} \right) |_{ \{ p_i \} \times {\cal M} } 
\right).
\end{displaymath}
These are, however, precisely the local sections of the sheaf
$\left( \alpha^* {\cal E} \right) |_{ \{ p_i \} \times {\cal M} }$ 
over the open set $U \subseteq {\cal M}$.
Thus, we have a map of local sections
\begin{displaymath}
{\cal F}(U) \: \longrightarrow \: \left( \alpha^* {\cal E} \right) |_{ \{ p_i
\} \times {\cal M} }(U)
\end{displaymath}
as desired, completing the description of the maps $\psi_i$.

Note that if ${\cal E}=TX$, then the composition
\[
T{\cal M}={\cal F}\to \alpha^*{\cal E}
|_{\{p_i\}\times{\cal M}}= ev_i^*{\cal E} = ev_i^*TX
\] 
is easily checked to be the differential of the evaluation map $ev_i:{\cal M}
\to X$.  Thus, in this case, the map 
$\psi_i: H^q\left( X, \Lambda^p {\cal E}^{\vee} \right) 
\mapsto H^q\left( {\cal M}, \Lambda^p {\cal F}^{\vee} \right)$ coincides
with the pullback by $ev_i$ on ordinary cohomology
\[
H^{p,q}(X)\to H^{p,q}({\cal M}).
\]
It follows that on the $(2,2)$ locus, the correlation functions coincide with
those of ordinary Gromov-Witten theory, with minor modifications due to
the turning off of topological gravity.

In summary, the heterotic chiral ring is a generalization of the
chiral ring of Gromov-Witten theory and quantum cohomology, and as such,
deserves to be better understood mathematically.  

In the next section, we relate to standard Gromov-Witten theory
a little more closely.

\subsection{Stable maps and compactifications} \label{stable}

In this section, we describe how to modify our basic construction of
the preceding sections to more realistic situations.  

Suppose we want
to compute a correlation function with $k$ insertions.  Let
$\M{g}{k}(X,d)$ be the moduli space of genus $g$ stable maps maps to
$X$ of degree $d$\footnote{More properly, we should replace $d$ with a
homology class $\beta\in H_2(X,{\bf Z})$, but we just denote this by $d$ and
call it a ``degree'' for ease of locution.} as introduced in \cite{kontenum}.  
These are maps $f:C\to
X$ of degree $d$, where $C$ is a connected algebraic curve of genus
$g$ with marked points $p_1,\ldots,p_n$ in the smooth locus of $C$
modulo automorphisms of $C$ preserving $f$ and the $p_i$.  The
stability condition is that each genus 0 component of $C$ on which $f$ is
constant has at most 3 special (nodal or marked) points.  
There are evaluation maps $e_i:\M{g}{k}(X,d)\to X$ defined by $e_i(f)=f(p_i)$.

There is a
forgetful map $\rho:\M{g}{k}(X,d)\to\M{g}{k}$ which forgets the map
and contracts any components of $C$ which have become unstable after
forgetting the map.  Let $\Sigma,p_1,\ldots,p_k\in\M{g}{n}$ be our worldsheet
with its fixed complex structure and insertion points.  For ease of notation
we denote this simply by $\Sigma\in\M{g}{n}$.  Then our more realistic model
for the instanton moduli space is
\[
{\cal M} = \rho^{-1}(\Sigma).
\]
Note that ${\cal M}$ is compact, and its elements correspond to degree $d$
maps $\phi:\Sigma\to X$, as well as maps from the union of $\Sigma$ with 
trees of 2~spheres to $X$ of total degree $d$.

Now there is still a universal map.  It is well known that the universal
family of curves over $\M{g}{k}(X,d)$ is given by $\M{g}{k+1}(X,d)$.  
There is a map 
$\pi_k:\M{g}{k+1}(X,d)\to \M{g}{k}(X,d)$ which is given by forgetting
the last marked point and contracting unstable components.  The fiber of
$\pi_k$ corresponds to the location of an extra point on $C$, which explains
why $\M{g}{k+1}(X,d)$ is the universal curve over $\M{g}{k}(X,d)$.  The map
$e_{k+1}:\M{g}{k+1}(X,d)\to X$ is then clearly the universal map.  There
are $k$ natural sections $s_i:\M{g}{k}(X,d)\to \M{g}{k+1}(X,d)$ of $\pi_k$ 
given by sending a stable map $f:C\to X$ with marked points $p_i$ to
the point $p_i$, identified with the corresponding point on the universal 
curve.

We now let ${\cal C}=\pi_k^{-1}({\cal M})$, and let $\alpha$ be the restriction
of $e_{k+1}$ to ${\cal C}$.  Then let $\pi$ be the restriction of
$\pi_k$ to ${\cal C}$.  This gives the diagram
\begin{displaymath}
\xymatrix{
{\cal C} \ar[r]^{\alpha} \ar[d]^{\pi} & X \\
{\cal M} & 
}.
\end{displaymath}
We now see that this is the proper
generalization of the universal map $\alpha:\Sigma\times{\cal M}\to X$ and
projection $\pi:\Sigma\times{\cal M}\to {\cal M}$ considered previously.

In particular, we can now define
\[
{\cal F}=\pi_*\alpha^*{\cal E}
\]
as before.  To get the generalized maps 
$\psi_i: H^q\left( X, \Lambda^p {\cal E}^{\vee} \right) 
\mapsto H^q\left( {\cal M}, \Lambda^p {\cal F}^{\vee} \right)$, we just
use the sections $s_i:\M{g}{k}(X,d)\to \M{g}{k+1}(X,d)$ in place of the
embeddings ${\cal M}=\{p_i\}\times{\cal M}
\subset\Sigma\times {\cal M}$ and proceed as before.

The price to be paid for this general construction using well-known
mathematics is that these ${\cal M}$ are almost never smooth except in 
simple cases like projective spaces and $g=0$.  So to evaluate correlation
functions, we would need to develop a virtual fundamental class in this
context.  This is very similar to Gromov-Witten theory, except that topological
gravity is turned off (by restricting to the fiber of $\rho$ and the 
virtual fundamental class must be modified accordingly).  Furthermore, the
techniques of Gromov-Witten theory do not apply in general.  However, the
techniques of localization can be applied if $X$ admits a torus
action and the bundle ${\cal E}$ is equivariant for that torus action.  Such
bundles are studied in \cite{klyachko,ericallen} 
if $X$ is a toric variety.  We do not
develop these techniques here since one of our goals in the present paper
is to verify the claims of \cite{abs} and the gauge bundles appearing there
are not of this type.   It would nevertheless be interesting to develop
computational techniques for this equivariant situation.

In Section~\ref{lsmmod} we will describe another compactification, the linear
sigma model compactification.  We expect this to coincide with the nonlinear
heterotic theory for simple spaces such as products of projective spaces,
and, by analogy with \cite{daveronen}, to be related to the heterotic theory
by a change of variables.

\subsection{Generalization of obstruction sheaves}   \label{excesszeromodes}

Next, let us turn to the case in which 
\begin{displaymath}
R^1 \pi_* \alpha^* TX \: \neq \: 0, \: \: \:
R^1 \pi_* \alpha^* {\cal E} \: \neq \: 0
\end{displaymath}
{\it i.e.} the case that would call for obstruction sheaves
on the $(2,2)$ locus.
Our analysis in section~\ref{quant:int} only makes sense
physically in the case that there are $\psi^i$ zero modes,
but no $\psi^{ \overline{\imath} }$ zero modes,
and $\lambda^{\overline{a}}$ zero modes, but no $\lambda^b$
zero modes.  The index theorems quoted in section~\ref{quant:int}
only specify the difference between the number of such zero modes.

We shall describe a proposal for how this case should be
described mathematically, that will generalize the obstruction
bundle story of the $(2,2)$ locus.  We shall check that our proposal satisfies
basic consistency tests, and reduces to the ordinary obstruction
bundle story on the $(2,2)$ locus.  However, there is a crucial
mathematical statement we have not yet been able to prove.

Write the number of zero modes of the four types of worldsheet
fermions as follows:
\begin{displaymath}
\begin{array}{rl}
\psi^i: & h^0\left( \Sigma, \phi^* TX \right) \: = \: m + p \\
\psi^{ \overline{\imath} }: & h^1\left( \Sigma, \phi^* TX \right) \: = \: p \\
\lambda^a: & h^1\left( \Sigma, \phi^* {\cal E} \right) \: = \: q \\
\lambda^{ \overline{a} }: & h^0\left( \Sigma, \phi^* {\cal E} \right) \: = \:
r \: + \: q
\end{array}
\end{displaymath}
The numbers $m$, $r$ are calculated by index theory,
and the numbers $p$, $q$ count `excess' zero modes.

We still need to make a simplifying assumption.  We assume that the
number of excess modes $p$ and $q$ are constant on the instanton moduli
space ${\cal M}$, although possibly nonzero.  As a consequence,
${\cal M}$ will still be smooth, of dimension $m+p$, and
${\cal F}$ will still be a bundle, of rank $r+q$.  But now the 
selection rule on the correlators gives bundle-valued forms of
degree $m$, $r$ -- so the selection rules no longer define
top forms on the moduli space in the presence of excess zero modes.
We will fix this problem by wedging with more bundle-valued differential
forms, but before we discuss the mathematics, let us review the physics.

Just as in \cite{pauldave}, we can soak up these excess zero modes
by using the four-fermi term in the worldsheet action.
Recall the four-fermi term has the form
\begin{equation}   \label{024fermi}
\int_{\Sigma} F_{i \overline{\jmath} a \overline{b} } \psi^i
\psi^{ \overline{\jmath} } \lambda^a \lambda^{ \overline{b} }
\end{equation}
where $F$ is the curvature of the bundle ${\cal E}$.
(Note, in particular, that the structure of the four-fermi term in
a sigma model is asymmetric between the left- and right-movers:
the curvature of the left-moving bundle ${\cal E}$ appears in the action,
but there is no term representing the curvature of the right-moving bundle
$TX$.)
Each time we bring down a factor of this four-fermi term,
we soak up one of each type of worldsheet fermion.
So long as $p=q$, which is the case for $(2,2)$ theories,
we can use this four-fermi term to soak up all of the excess zero modes.

In general, however, $p$ need not equal $q$.
Suppose, without loss of generality, that $p>q$.
Then we could use $q$ factors of the four-fermi terms to soak up
all of the excess $\lambda$ zero modes, and all but $p-q$ of the
excess $\psi$ zero modes.  We could bring down an additional
$p-q$ factors of the four-fermi term, and then use $\lambda$ propagators
to contract away the excess $\lambda$ fermions.  The resulting
correlation function would exhibit a dependence on the positions
on the worldsheet in which the correlators were inserted.
The result is not topological, but since this theory is not a topological
field theory, there is no good reason to believe that all
correlators under consideration can be expressed purely topologically.

As we are only interested in those correlators that can be expressed
purely topologically, henceforth we shall only consider theories for which
\begin{equation}   \label{constraint:balance}
H^1\left(\Sigma, \phi^* TX \right) \: = \:
H^1\left( \Sigma, \phi^* {\cal E} \right)
\end{equation}
for all $\phi$,
{\it i.e.} $p=q$ everywhere on ${\cal M}$.
Phrased more formally, we shall assume that
\begin{displaymath}
\mbox{rank } R^1 \pi_{2 *} \alpha^* {\cal E}
\: = \:
\mbox{rank }R^1 \pi_{2 *} \alpha^* TX
\end{displaymath}
everywhere on ${\cal M}$, i.e.\ that $p=q$ in the notation introduced above.

Let us now return to the mathematical description of our correlators.
As outlined above, the physical selection rule says that images
of the vertex operators as sheaf cohomology on ${\cal M}$ wedge 
together to form an element of
\begin{displaymath}
H^m\left( {\cal M}, \Lambda^r {\cal F}^{\vee} \right)
\end{displaymath}
but the dimension of ${\cal M}$ is $m+p$, and the rank of ${\cal F}$ is
$r+q$, so we need an additional factor if we wish the
correlation functions to be expressible as integrals of
top forms.  

This problem has a well-known solution on the $(2,2)$ locus.
In \cite{pauldave}, bringing down factors of the four-fermi term
was interpreted in terms of wedging the differential
forms representing correlators (whose total degree was too small
to be a top form)
with a differential form representing the Euler class
(coming from the four-fermi term factors), which had
the effect of making the integrand a top form, so that
correlation functions again naturally generate numbers.  

There is an analogous phenomenon here.  If we try to interpret each
$(0,2)$ four-fermi term~(\ref{024fermi}) as a bundle-valued differential
form on the moduli space, then each such four-fermi term should be
identified with an element of
\begin{equation}    \label{4fermiinterp}
H^1\left( {\cal M}, {\cal F}^{\vee} \otimes {\cal F}_1 \otimes
{\cal G}_1^{\vee} \right)
\end{equation}
on symmetry grounds, where
\begin{eqnarray*}
{\cal F}_1 & \equiv & R^1 \pi_{*} \alpha^* {\cal E}, \\
{\cal G}_i & \equiv & R^i \pi_{*} \alpha^* TX.
\end{eqnarray*}
are the sheaves over the moduli space defined by the zero modes
of the fermions.
(The $\psi^i$ is responsible for having degree
one sheaf cohomology; the other three fermions are responsible for the
coefficients.)

We should stress at this point, however, that this is an ansatz.
One would like to be able to prove mathematically that the curvature
$F$ of ${\cal E}$ defines an element of the sheaf cohomology group
above, but we have not yet been able to do this.

Nevertheless, we can check that this ansatz is consistent.
First, we shall show that with this ansatz, correlation functions
can be expressed as integrals of top forms, {\it i.e.} naturally
generate numbers.
Previously in section~\ref{quant:int}, when
$R^1 \pi_{*} \alpha^* {\cal E}$ and $R^1 \pi_{*} \alpha^* TX$
both vanished, we saw that the heterotic anomaly cancellation condition
and Grothendieck-Riemann-Roch implied that
\begin{displaymath}
\Lambda^{top} {\cal F} \: = \: K_{ {\cal M} }^{\vee}
\end{displaymath}
exactly as needed for our correlation functions to generate a number.
However, more generally we have a slightly different statement.
If we define
\begin{eqnarray*}
{\cal F}_1 & \equiv & R^1 \pi_{*} \alpha^* {\cal E} \\
{\cal G}_i & \equiv & R^i \pi_{*} \alpha^* TX,
\end{eqnarray*}
as above,
then the anomaly cancellation condition and Grothendieck-Riemann-Roch imply
\begin{displaymath}
\Lambda^{top} {\cal F} \otimes \Lambda^{top} {\cal F}_1^{\vee}
\: \cong \: 
\Lambda^{top} {\cal G}_0 \otimes \Lambda^{top} {\cal G}_1^{\vee}.
\end{displaymath}

In particular, since ${\cal G}_0 \cong T {\cal M}$,
this means that
\begin{displaymath}
\Lambda^{top} {\cal F}^{\vee} \otimes
\Lambda^{top} {\cal F}_1 \otimes \Lambda^{top} {\cal G}_1^{\vee}
\: \cong \:
K_{ {\cal M} }
\end{displaymath}
so that we can fix up correlation functions by wedging with a representative
of
\begin{equation}   \label{fixfactor}
H^{q}\left( {\cal M}, \Lambda^q {\cal F}^{\vee} \otimes
\Lambda^{q} {\cal F}_1 \otimes \Lambda^{q} {\cal G}_1^{\vee} \right)
\end{equation}
(recall ${\cal F}_1$ has rank $q$, matching the rank of ${\cal G}_1$).
Note, however, that if we bring down enough copies of the
$(0,2)$ four-fermi term to absorb the `excess' zero modes,
then from our ansatz~(\ref{4fermiinterp}), we generate
precisely the factor~(\ref{fixfactor}) above.
In other words, our interpretation of the $(0,2)$ four-fermi terms
is precisely what we need to describe correlators as integrals
of top forms.

Let us now check that the description above gives correct
results on the $(2,2)$ locus.  In this case, ${\cal F}_1\simeq
{\cal G}_1$ so that the ${\cal F}_1$ and ${\cal G}_1$ factors in 
(\ref{fixfactor}) cancel out.  Nevertheless, the particular class
of (\ref{fixfactor}) used depends on ${\cal G}_1$, as we will see.

Recall that in the $(2,2)$
case, correlators are described by differential forms on ${\cal M}$,
not sheaf cohomology groups, and the factor~(\ref{fixfactor}) is 
replaced by the
top Chern class of ${\cal G}_1$, {\it i.e.} $c_q\left( {\cal G}_1 \right)$.

Chern classes and sheaf cohomology can be related via the Atiyah class
of the bundle, which we shall briefly review.
Consider expressing Chern classes in terms of the curvature
of the connection on a holomorphic vector bundle ${\cal E}$
on a K\"ahler manifold $X$:
\begin{displaymath}
\mbox{c}_r \: \propto \: \mbox{Tr } F \wedge F \wedge \cdots \wedge F.
\end{displaymath}
Taking advantage of hermitian fiber metrics on vector bundles,
we can write 
\begin{displaymath}
F \: = \: F_{i \overline{\jmath} a \overline{b} }
dz^i \wedge d \overline{z}^{ \overline{\jmath} } \wedge
\lambda^a \wedge \lambda^{ \overline{b} }.
\end{displaymath}
Furthermore, because of the Bianchi identity,
if $F$ is a holomorphic connection ({\it i.e.} $F_{i j} = F_{
\overline{\imath} \overline{\jmath} } = 0$), then 
$F$ is $\overline{\partial}$-closed.
Thus, we can think of $F$ as a holomorphic $(0,1)$-form valued in
$TX^{\vee} \otimes {\cal E} \otimes {\cal E}^{\vee}$,
and so in particular the curvature $F$ defines an element of
\begin{displaymath}
H^1\left(X, TX^{\vee} \otimes {\cal E} \otimes {\cal E}^{\vee} \right).
\end{displaymath}
The sheaf cohomology class above is known as the Atiyah
class and is independent of the choice of connection \cite{atiyah}.
Taking the trace defines a map
\begin{displaymath}
H^1\left(X, TX^{\vee} \otimes {\cal E} \otimes {\cal E}^{\vee} \right)
\: \mapsto \:
H^1\left(X, TX^{\vee} \right) \: \cong \: H^{1,1}(X)
\end{displaymath}
whose image is the first Chern class of the bundle.
Furthermore, note that by wedging $r$ copies of $F$ together we create
an element of
\begin{displaymath}
H^r\left(X, \Lambda^r TX^{\vee} \otimes \Lambda^r {\cal E} \otimes
\Lambda^r {\cal E}^{\vee} \right)
\end{displaymath}
and the trace defines a map from the sheaf cohomology
group above to
\begin{displaymath}
H^r\left(X, \Lambda^r TX^{\vee} \right) \: \cong \: H^{r,r}(X)
\end{displaymath}
whose image is the $r$th Chern class.

In particular, on the $(2,2)$ locus, the factor~(\ref{fixfactor})
is the sheaf cohomology description of the $q$th Chern class of
${\cal G}_1$, as claimed.

Thus, even though we do not yet understand mathematically 
how the Atiyah class of ${\cal E}$
induces an element of the sheaf cohomology group~(\ref{4fermiinterp}),
our proposal satisfies reasonable consistency tests.
This gives us some confidence in our interpretation.

\section{Linear-sigma-model-based compactifications}  \label{lsmmod}

\subsection{Generalities}

Linear sigma models were used in \cite{daveronen} to provide natural
compactifications of moduli spaces of worldsheet instantons.
Let us take a moment to review their construction.
Recall that given a toric variety expressed as a GIT quotient
of a set of homogeneous polynomials $x_a$ by a set of ${\bf C}^{\times}$
actions, 
the linear sigma model moduli space
is described as a GIT quotient of the space of zero modes of the $x_a$.
Let $\vec{n}_a$ denote the weights of homogeneous coordinate $x_a$
with respect to the ${\bf C}^{\times}$ actions, and let
$\vec{d}$ denote the (fixed) degree of the worldsheet instantons
(expressed as a vector of weights with respect to the same
${\bf C}^{\times}$ actions). Then we can expand each $x_a$ as
\begin{eqnarray*}
x_a & \in & H^0\left( {\bf P}^1, {\cal O}(\vec{n}_a \cdot \vec{d} ) \right) \\
& = & x_{a0} u^{d_a} \: + \: x_{a1} u^{d_a-1} v \: + \: \cdots
\end{eqnarray*}
for $d_a = \vec{n}_a \cdot \vec{d}$, where $u$, $v$ are homogeneous
coordinates on the worldsheet ${\bf P}^1$.
To construct the linear sigma model moduli space,
we take the space of $x_{ai}$, and 
quotient by the same set of ${\bf C}^{\times}$'s
as defined the original toric variety, such that each $x_{ai}$ has
weight vector $\vec{n}_a$ (same as for the original $x_a$),
after removing an exceptional set.

The original (2,2) gauged linear sigma models of \cite{phases}
were generalized in \cite{distkachru} to describe (0,2) theories,
that is, Calabi-Yau's together with bundles.
The bundles are presented physically in a very special form,
as the cohomology of a short complex
\begin{displaymath}
\oplus {\cal O} \: \stackrel{E}{\longrightarrow} \: 
\oplus {\cal O}(\vec{n}_a) \:
\stackrel{F}{\longrightarrow} \: \oplus {\cal O}(\vec{m}_i)
\end{displaymath}
sometimes known as a `monad'.  The bundles in question are bundles
over\footnote{In this paper, we will work with monads describing bundles
over the ambient space.  However, historically monads have been used
to describe bundles only over a Calabi-Yau complete intersection.
The distinction involves whether the composition of maps $F \circ E$ vanishes
identically (as we will assume throughout this paper) or only
up to hypersurface equations (as is more typical in the older
literature \cite{dgm,distkachru}).} the ambient toric variety. If a superpotential is present to
specify a Calabi-Yau subvariety, then the bundle on the subvariety
is obtained by restricting the bundle on the ambient space.

The methods of \cite{daveronen} can be extended in a very straightforward
fashion to define extensions of the sheaves ${\cal F}$, ${\cal F}_1$
over the (compact) linear sigma model moduli spaces, as we will discuss
in numerous examples in the rest of this section.
The basic idea is to expand each worldsheet field corresponding to a line
bundle, in a basis of zero modes, and associate the coefficients
to line bundles on ${\cal M}$.  For example, if we had a reducible
bundle ${\cal E} = \oplus_a {\cal O}(\vec{n}_a)$, described by 
a set of free left-moving fermions of charges $\vec{n}_a$,
then our ansatz yields the induced bundles
\begin{eqnarray*}
{\cal F} & = & \oplus_a H^0\left( {\bf P}^1, {\cal O}(
\vec{n}_a \cdot \vec{d} ) \right) \otimes_{ {\bf C} }
{\cal O}(\vec{n}_a), \\
{\cal F}_1 & = & \oplus_a H^1\left( {\bf P}^1,
{\cal O}(\vec{n}_a \cdot \vec{d} ) \right) \otimes_{ {\bf C} }
{\cal O}(\vec{n}_a).
\end{eqnarray*}
It can be shown\footnote{Clearly 
\begin{displaymath}
R^{i} \pi_* \alpha^* {\cal E} \: = \: \oplus_a R^{i} \pi_* \alpha^*
{\cal O}(\vec{n}_a)
\end{displaymath}
for $i=0,1$, and these sheaves 
have the same ranks as the ${\cal F}$, ${\cal F}_1$ listed.
To show that the $R^i \pi_* \alpha^* {\cal E}$ completely decompose
into a direct sum of line bundles one uses the $T$-equivariant nature
of the line bundles ${\cal O}(\vec{n}_a)$, as we will describe in detail
in the section on bundles presented as cokernels.}
that this ansatz matches the $R^i \pi_* \alpha^* {\cal E}$
on the open subset of ${\cal M}$ corresponding to honest maps,
and furthermore that this ansatz has the desired rank for ${\cal F}
\ominus {\cal F}_1$ as well as the correct determinant line bundle 
$\Lambda^{top} {\cal F}^{\vee} \otimes
\Lambda^{top} {\cal F}_1$.  These verifications are special cases of
arguments we will present later, so we omit details.

We will describe more general cases explicitly in the rest of this
section.
We will also check that this ansatz for induced sheaves
is compatible with
needed results, {\it e.g.} that
\begin{displaymath}
\Lambda^{top} {\cal F}^{\vee} \otimes \Lambda^{top} {\cal F}_1 \: \cong \:
K_{ {\cal M} } \otimes \Lambda^{top} {\cal G}_1 
\end{displaymath}
continues to hold after extending the sheaves over the compactification
of the moduli space, and that in all cases, our ansatz gives
sheaves that agree with the $R^{i} \pi_* \alpha^* {\cal E}$ on the open
subset of ${\cal M}$ corresponding to honest maps.

An important technical issue is that the anomaly cancellation
condition in gauged linear sigma models is slightly
stronger than the mathematical statement 
$\mbox{ch}_2({\cal E}) = \mbox{ch}_2(TX)$, 
as shown in \cite{dgm}, and can even
distinguish different presentations of the
same gauge bundle.  We specifically require the stronger form,
as described in \cite{dgm}, in order for our constructions to be
consistent.  In particular, in section~\ref{multpres} we shall see
some examples which satisfy $\mbox{ch}_2({\cal E}) = \mbox{ch}_2(TX)$
but fail the stronger linear sigma model anomaly cancellation condition,
despite merely being alternative presentations of well-behaved
linear sigma models.
In these examples, our construction fails.
We will discuss the linear sigma model anomaly cancellation condition
in more detail later.

A related issue, also discussed in section~\ref{multpres}, that the extension
of the sheaves  $R^{i} \pi_* \alpha^*
{\cal E}$ over the compactification divisor depends upon the presentation
of ${\cal E}$, and different presentations can give very different
extensions.

\subsection{Bundles presented as cokernels, and a check of the $(2,2)$ locus}

\subsubsection{Easy example:  tangent bundle of ${\bf P}^{N-1}$}

Let us begin with an easy example:  the tangent bundle of a projective
space ${\bf P}^{N-1}$,
which physically in the linear sigma model arises as a cokernel:
\begin{displaymath}
0 \: \longrightarrow \: {\cal O} \: \longrightarrow \:
\bigoplus_1^N {\cal O}(1) \: \longrightarrow \:
T {\bf P}^{N-1} \: \longrightarrow \: 0.
\end{displaymath}
Physically both the ${\cal O}$ and the $\oplus {\cal O}(1)$
correspond to worldsheet fields, the latter to $N$ (0,2) fermi multiplets
of charge one, and the former to a single neutral (0,2) chiral multiplet.

Thus, when we consider a moduli space of degree $d$ maps
${\bf P}^1 \rightarrow {\bf P}^{N-1}$,
we want to replace the worldsheet fields by their zero modes,
coupling to bundles on the moduli space determined by the scaling
properties of the original linear sigma model fields with respect
to the toric action.  Furthermore, those zero modes should obey relations
determined by the relations in the original linear sigma model.

In other words, on the moduli space, the sheaf ${\cal F}$ on the moduli space
corresponding to $T {\bf P}^{N-1}$ is given by
\begin{eqnarray*}
\lefteqn{
\mbox{coker } \left\{ \:
H^0\left( {\bf P}^1, {\cal O}(d \cdot 0) \right) \otimes_{ {\bf C} }
{\cal O} \: \longrightarrow \:
\bigoplus_1^N H^0\left( {\bf P}^1, {\cal O}(d \cdot 1) \right) \otimes_{ {\bf C} }
{\cal O}(1) \:
\right\} 
} \\
& \hspace*{1.5in}  = & 
\mbox{coker } \left\{ \:
{\cal O} \: \longrightarrow \: 
\bigoplus_1^{N(d+1)} {\cal O}(1) 
\: \right\}
\end{eqnarray*}
where the maps are given by multiplication by homogeneous coordinates
on the moduli space.

Let us take a moment to work through the maps in more detail.
If the homogeneous coordinates on ${\bf P}^{N-1}$ are denoted $x_i$,
and homogeneous coordinates on the worldsheet are denoted $u$, $v$,
then for degree $d$ maps we can write
\begin{displaymath}
x_i \: = \: c_{i0} u^d \: + \: c_{i1} u^{d-1}v \: + \: \cdots \: + \:
c_{id} v^d.
\end{displaymath}
Similarly, arbitrary local sections of ${\cal O}(1)$ can be written as
\begin{displaymath}
\lambda_i \: = \: \omega_{i0} u^d \: + \: \omega_{i1} u^{d-1} v
\: + \: \cdots \: + \: \omega_{id} v^d.
\end{displaymath}
In the short exact sequence defining ${\cal F}$,
the map ${\cal O} \rightarrow {\cal O}(1)^{N(d+1)}$ arising
from the original map ${\cal O}\to {\cal O}(1)^N$ on ${\bf P}^{N-1}$
that multiplied
local sections by $x_i$, can now be expressed in terms of the $u^i v^j$.
In other words, at the level of zero modes,
that map now multiplies $1$ by $c_{ik}$ to land in the bundle
whose sections are generated by $\omega_{ik}$.
More generally, any map ${\cal O}(n) \rightarrow {\cal O}(m)$
on the original space induces a map
\begin{equation}
\label{inducedmap}
H^0\left( {\bf P}^1, {\cal O}(nd) \right) \otimes_{ {\bf C} }
{\cal O}(n)
\: \longrightarrow \:
H^0\left( {\bf P}^1, {\cal O}(md) \right) \otimes_{ {\bf C} }
{\cal O}(m)
\end{equation}
on the moduli space, given by expressing the original map
(evaluated on zero modes) in terms of its $u^i v^j$ components.
More explicitly, the original map is given by 
multiplication by a polynomial
of degree $m-n$, which pulls back to a polynomial of degree $d(m-n)$ on
${\bf P}^1$ whose coefficients depend on the $c_{ij}$, and which in addition 
has degree $m-n$ in the $c_{ij}$.  This is expressed in more invariant
fashion as a map (\ref{inducedmap}).

Since the $c_{ik}$ are homogeneous coordinates on the moduli space,
it should now be clear that
this sheaf is $T {\bf P}^{N(d+1)-1} = T {\cal M}$,
exactly as desired.

\subsubsection{More general bundles presented as cokernels}

More generally, given a bundle ${\cal E}$ presented in the linear
sigma model as a cokernel of the form
\begin{displaymath}
0 \: \longrightarrow \: {\cal O}^{\oplus k} \: \longrightarrow \:
\oplus_a {\cal O}(\vec{n}_a) \: \longrightarrow \: {\cal E}
\: \longrightarrow \: 0
\end{displaymath}
for worldsheet instantons of degree $d$ we have an induced long
exact sequence
\begin{displaymath}
\begin{array}{ccccccc}
0 & \longrightarrow & \oplus_k H^0\left( {\bf P}^1, 
{\cal O}(\vec{0} \cdot \vec{d})
\right)
\otimes_{ {\bf C} } {\cal O} & \longrightarrow &
\oplus_a H^0\left( {\bf P}^1, {\cal O}(
\vec{n}_a \cdot \vec{d}) \right) \otimes_{ {\bf C} } {\cal O}(\vec{n}_a) & 
\longrightarrow & {\cal F} \\ 
& \longrightarrow & \oplus_k H^1\left( {\bf P}^1, {\cal O}(\vec{0} \cdot
\vec{d} ) \right) \otimes_{ {\bf C} } {\cal O} & \longrightarrow &
\oplus_a H^1\left({\bf P}^1, {\cal O}(\vec{n}_a \cdot \vec{d}) \right) 
\otimes_{ {\bf C} } {\cal O}(
\vec{n}_a ) & \longrightarrow & {\cal F}_1 \\
 & \longrightarrow & 0 & & & & 
\end{array}
\end{displaymath}
This simplifies to give the short exact sequence,
\begin{equation}
0 \: \longrightarrow \: {\cal O}^{\oplus k} \: \longrightarrow \:
\oplus_a H^0\left( {\bf P}^1, {\cal O}(\vec{n}_a \cdot \vec{d} ) \right)
\otimes_{ {\bf C} } {\cal O}(\vec{n}_a ) \: \longrightarrow \: {\cal F} \:
\longrightarrow \: 0
\end{equation}
and the relation
\begin{equation}    \label{f1split}
{\cal F}_1 \: \cong \:
\oplus_a H^1\left( {\bf P}^1, {\cal O}( \vec{n}_a \cdot \vec{d} )\right)
\otimes_{ {\bf C} } {\cal O}(\vec{n}_a).
\end{equation}

To verify that this method is giving reasonable results,
we shall next check the following things:
\begin{itemize}
\item First, we shall check that this gives the
correct results on the $(2,2)$ locus.
\item We shall argue that more generally, 
${\cal F}$ and ${\cal F}_1$ agree with $R^0 \pi_* \alpha^* {\cal E}$,
$R^1 \pi_* \alpha^* {\cal E}$, respectively, on the open subset of
${\cal M}$ corresponding to honest maps.
\item Finally, we shall check that $\mbox{rank } \left(
{\cal F} \ominus {\cal F}_1 \right)$ and $\Lambda^{top} {\cal F}^{\vee}
\otimes \Lambda^{top} {\cal F}_1$ satisfy the correct relations for our
formal analysis of correlation functions to proceed.
\end{itemize}

First, let us check that our results are correct on the $(2,2)$ locus.
Note that when ${\cal E}$ is the tangent bundle of a toric variety,
${\cal F}$ as described above is the tangent bundle to the linear
sigma moduli space of \cite{daveronen}.

In the case of the tangent bundle to ${\bf P}^{N-1}$ above,
$\vec{n}_a \cdot \vec{d} = d$ for all $a$, so as long as $d > 0$,
we have that
${\cal F}_1 = 0$ for the tangent bundle.  

When ${\cal E} = TX$, the sheaf ${\cal F}_1$ is known as the obstruction
bundle\footnote{Technically, we shall show that ${\cal F}_1$
agrees with the obstruction sheaf over the open subset of the
moduli space corresponding to honest maps.  Strictly speaking,
we are not aware of a previous general definition of obstruction sheaf over 
(compact) linear
sigma model moduli spaces.}, 
as we shall now check.
First, a couple of easy tests of this statement. 
Loosely, the obstruction sheaf is the sheaf over the moduli space
defined by the $\psi_+^{\overline{\imath}}$ zero modes, and that is 
precisely the physical meaning of ${\cal F}_1$.
More concretely,
since 
\begin{displaymath}
h^1\left( {\bf P}^1, {\cal O}( \vec{n}_a \cdot \vec{d} ) \right) \: = \:
\left\{ \begin{array}{cl}
        - \vec{n}_a \cdot \vec{d} -1 & \vec{n}_a \cdot \vec{d} \leq -1 \\
         0 & \mbox{otherwise}
        \end{array} \right.
\end{displaymath}
we see that the top Chern class of the obstruction sheaf
${\cal F}_1$ is given by 
\begin{displaymath}
\prod_{ \vec{n}_a \cdot \vec{d} \leq -1 } 
c_1( {\cal O}(\vec{n}_a) )^{ - \vec{n}_a \cdot \vec{d} -1 },
\end{displaymath}
a result that precisely matches~\cite{daveronen}[equ'n (3.62)].

Next, let us check more systematically that on the open subset of
the linear sigma model moduli space ${\cal M}$ corresponding to
honest maps, ${\cal F}_1$ really is the obstruction sheaf.
Recall that on the (2,2) locus, the obstruction sheaf is
$R^1 \pi_* \alpha^* TX$, at least over the locus where there is a universal
instanton $\alpha: {\bf P}^1 \times {\cal M} \rightarrow X$.  Now if $X$ is
toric, present the tangent bundle as
\begin{displaymath}
0 \: \longrightarrow \: {\cal O}^{\oplus m} \: \longrightarrow \:
\oplus_a {\cal O}(\vec{n}_a ) \: \longrightarrow \: TX \:
\longrightarrow \: 0.
\end{displaymath}
Applying $\pi_* \alpha^*$ gives rise to an exact sequence including terms
\begin{displaymath}
R^1\pi_* \alpha^* {\cal O}^{\oplus m} \: \longrightarrow \:
\oplus_a R^1 \pi_* \alpha^* {\cal O}(\vec{n}_a) \: \longrightarrow \:
R^1 \pi_* \alpha^* TX \: \longrightarrow \: 0.
\end{displaymath}
The first term is zero since $H^1({\bf P}^1, {\cal O}) = 0$, so we get an
isomorphism
\begin{displaymath}
R^1 \pi_* \alpha^* TX \: \cong \: \oplus_a R^1 \pi_* \alpha^*
{\cal O}(\vec{n}_a).
\end{displaymath}

To reconcile with the computation of ${\cal F}_1$ in
equation~(\ref{f1split}), 
we need to compare $R^1 \pi_* \alpha^* {\cal O}(\vec{n}_a)$ with
\begin{displaymath}
H^1\left( {\bf P}^1, {\cal O}(\vec{n}_a \cdot \vec{d}) \right)
\otimes_{ {\bf C} } {\cal O}(\vec{n}_a).
\end{displaymath}
These sheaves both have the
same ranks, since $H^1\left( {\bf P}^1, {\cal O}(\vec{n}_a \cdot \vec{d})
\right)$ is the fiber of $R^1 \pi_* \alpha^* {\cal O}(\vec{n}_a)$.
In addition, note that the induced torus action decomposes
$R^1 \pi_* \alpha^* {\cal O}(\vec{n}_a)$ into eigenbundles, each of which
have precisely the same torus weight as the bundle ${\cal O}(\vec{n}_a)$.
Since there is a single weight and ${\cal M}$ is itself a toric variety,
we simply get a sum of line bundles, as expected.
Thus, our sheaf ${\cal F}_1$ really does match the obstruction sheaf
on the (2,2) locus, as advertised.

We shall outline an example of an obstruction sheaf in section~\ref{obs}.
Note that from equation~(\ref{f1split}), we see that
the obstruction sheaf over any linear sigma model moduli space is always
a direct sum of line bundles.

Now that we have checked that on the $(2,2)$ locus
the sheaves ${\cal F}$, ${\cal F}_1$ are precisely
the tangent bundle and obstruction sheaf, as expected,
next we shall check that ${\cal F}$ and ${\cal F}_1$
match $R^0 \pi_* \alpha^* {\cal E}$ and
$R^1 \pi_* \alpha^* {\cal E}$ on the open subset of ${\cal M}$ 
corresponding to honest maps.  The analysis is very similar to our
earlier comparison of ${\cal F}_1$ on the $(2,2)$ locus to the
obstruction sheaf.  To do this, we return to the definition
of ${\cal E}$ as a cokernel
\begin{displaymath}
0 \: \longrightarrow \: {\cal O}^{\oplus k} \: \longrightarrow \:
\oplus_a {\cal O}(\vec{n}_a) \: \longrightarrow \:
{\cal E} \: \longrightarrow \: 0
\end{displaymath}
which induces, on the open subset of ${\cal M}$ corresponding to honest
maps, the long exact sequence
\begin{displaymath}
\begin{array}{ccccccc}
0 & \longrightarrow & \oplus_k R^0 \pi_* \alpha^* {\cal O} & \longrightarrow
& \oplus_a R^0 \pi_* \alpha^* {\cal O}(\vec{n}_a) & \longrightarrow &
R^0 \pi_* \alpha^* {\cal E} \\
& \longrightarrow & \oplus_k R^1 \pi_* \alpha^* {\cal O} & \longrightarrow
& \oplus_a R^1 \pi_* \alpha^* {\cal O}(\vec{n}_a) &
\longrightarrow & R^1 \pi_* \alpha^* {\cal E} \\
& \longrightarrow & 0. & & & & 
\end{array}
\end{displaymath}
Since $R^1\pi_* \alpha^* {\cal O} = 0$, this long exact sequence simplifies
to become the short exact sequence,
\begin{displaymath}
0 \: \longrightarrow \: {\cal O}^{\oplus k} \: \longrightarrow \:
\oplus_a R^0\pi_* \alpha^* {\cal O}(\vec{n}_a) \: \longrightarrow \:
R^0 \pi_* \alpha^* {\cal E} \: \longrightarrow \: 0
\end{displaymath}
and the relation
\begin{displaymath}
R^1 \pi_* \alpha^* {\cal E} \: \cong \:
\oplus_a R^1 \pi_* \alpha^* {\cal O}(\vec{n}_a).
\end{displaymath}
For the same reasons as in our discussion of the obstruction sheaf,
the sheaves $R^{i} \pi_* \alpha^* {\cal O}(\vec{n}_a)$ all split
into a direct sum of line bundles, and so we now see explicitly
that when restricted
to the open subset $U$ of ${\cal M}$ corresponding to honest maps,
${\cal F}|_U \cong R^0 \pi_* \alpha^* {\cal E}$
and ${\cal F}_1|_U \cong R^1 \pi_* \alpha^* {\cal E}$.

Let us next check that the
sheaves ${\cal F}$, ${\cal F}_1$ derived above do indeed possess the properties
discussed in section~\ref{excesszeromodes}, namely that
\begin{displaymath}
\Lambda^{top} {\cal F}^{\vee} \otimes
\Lambda^{top} {\cal F}_1 \: \cong \:
K_{ {\cal M} } \otimes \Lambda^{top} {\cal G}_1
\end{displaymath}
and that 
\begin{displaymath}
\mbox{rank } {\cal F} \: - \: \mbox{rank } {\cal F}_1 \: = \:
\chi( \phi^* {\cal E} ) \: = \: c_1(\phi^* {\cal E}) \: + \:
\mbox{rank } {\cal E}.
\end{displaymath} 

To perform this verification, let us present the tangent bundle to the
toric variety in the form
\begin{displaymath}
0 \: \longrightarrow \: {\cal O}^{\oplus m} \: \longrightarrow \:
\oplus_n {\cal O}(\vec{q}_n ) \: \longrightarrow \: TX \:
\longrightarrow \: 0.
\end{displaymath}
Writing $\vec{n} = (n^i)$, and using ${\cal E}$ as described above,
the anomaly cancellation condition in the (0,2) gauged linear sigma
model takes the form
\begin{displaymath}
\sum_a n_a^i n_a^j \: = \: \sum_n q_n^i q_n^j
\end{displaymath}
for each $i$, $j$, a condition which is very slightly stronger than the 
statement of matching second Chern characters \cite{dgm}.
Similarly, we shall also impose the requirement that for each $i$,
\begin{displaymath}
\sum_a n_a^i \: = \: \sum_n q_n^i,
\end{displaymath}
which is a slightly stronger form of the constraint
\begin{displaymath}
\Lambda^{top} {\cal E}^{\vee} \: \cong \: K_X.
\end{displaymath}

First, consider the ranks of ${\cal F}$ and ${\cal F}_1$.
\begin{eqnarray*}
\mbox{rank }{\cal F} & = & \sum_{\vec{n}_a \cdot \vec{d} \geq 0}
\left( \vec{n}_a \cdot \vec{d} \: + \: 1 \right) \: - \: k  \\
\mbox{rank } {\cal F}_1 & = & \sum_{ \vec{n}_a \cdot \vec{d} < 0 }
\left( - \vec{n}_a \cdot \vec{d} \: - \: 1 \right) 
\end{eqnarray*}
from which we see that
\begin{eqnarray*}
\mbox{rank } {\cal F} \: - \: \mbox{rank } {\cal F}_1 & = &
\mbox{rank } {\cal E} \: + \: \sum_a \vec{n}_a \cdot \vec{d} \\
& = & \mbox{rank } {\cal E} \: + \: c_1(\phi^* {\cal E})
\end{eqnarray*}
precisely as desired.

Next, consider the product of top exterior powers of ${\cal F}$
and ${\cal F}_1$.
{}From their expressions above, we see that
\begin{eqnarray*}
c_1({\cal F}) & = & \sum_{\vec{n}_a \cdot \vec{d} \geq 0}
\left( \vec{n}_a \cdot \vec{d} \: + \: 1 \right) n_a^i J_i \\
c_1({\cal F}_1) & = & \sum_{ \vec{n}_a \cdot \vec{d} < 0 }
\left(  - \vec{n}_a \cdot \vec{d} \: - \: 1 \right) n_a^i J_i 
\end{eqnarray*}
where the $J_i$ generate degree two cohomology of ${\cal M}$
(and may be constrained by relations).
Thus,
\begin{displaymath}
c_1\left( {\cal F} \ominus {\cal F}_1 \right) \: = \:
\left( \sum_a n_a^i J_i \right) \: + \: \left( \sum_a 
\left( \vec{n}_a \cdot \vec{d} \right) n_a^i J_i \right),
\end{displaymath}
where $\ominus$ denotes the K-theoretic difference.
For the tangent bundle,
\begin{displaymath}
c_1\left( T {\cal M} \ominus {\cal G}_1 \right) \: = \:
\left( \sum_n q_n^i J_i \right) \: + \:
\left( \sum_n \left( \vec{q}_n \cdot \vec{d} \right) q_n^i J_i \right).
\end{displaymath}
However, using the anomaly cancellation condition above plus the
statement that $\Lambda^{top} {\cal E}^{\vee} \: \cong \: K_X$,
we see immediately that
\begin{displaymath}
c_1\left( {\cal F} \ominus {\cal F}_1 \right) \: = \:
c_1\left( T {\cal M} \ominus {\cal G}_1 \right)
\end{displaymath}
precisely as desired.

\subsubsection{Example:  Obstruction sheaves and Hirzebruch surfaces}   
\label{obs}

So far we have proven general properties of this class of
$(0,2)$ gauged linear sigma models, and described details
of the special case of the tangent bundle of ${\bf P}^{N-1}$.
Next, we shall consider a special case on the $(2,2)$ locus
in which the obstruction sheaf is nontrivial, to further illustrate
these ideas.

Specifically, we will consider
 moduli spaces of maps into
the curve of self-intersection $-n$ on a Hirzebruch surface
${\bf F}_n$.  We can describe the Hirzebruch surface in terms of
four homogeneous coordinates $s$, $t$, $u$, $v$, with weights under
two ${\bf C}^{\times}$ actions as follows:
\begin{center}
\begin{tabular}{c|cccc} 
 & $s$ & $t$ & $u$ & $v$ \\ \hline
$\lambda$ & $1$ & $1$ & $n$ & $0$ \\
$\mu$ & $0$ & $0$ & $1$ & $1$ \\
\end{tabular}
\end{center}
The homogeneous coordinates on the ${\bf P}^1$ fibers are $u$, $v$.
Maps into the ${\bf P}^1$ fiber have weight $(0,1)$ under $(\lambda, \mu)$,
and similarly maps into the base have weight $(1,0)$.
Following \cite{beauville}[chapter IV], 
maps into the curve of self-intersection $-n$ should have weight
\begin{displaymath}
(1,0) \: - \: n (0,1) \: = \: (1,-n)
\end{displaymath}
under $(\lambda, \mu)$.
For such maps, when we expand in zero modes, we find that
$s$ and $t$ are sections of ${\cal O}_{ {\bf P}^1 }(1)$,
$u$ is a section of ${\cal O}_{ {\bf P}^1 }$,
and $v$ is a section of ${\cal O}_{ {\bf P}^1 }(-n)$.
Let $x$ and $y$ be homogeneous coordinates on the worldsheet
${\bf P}^1$, then we can expand
\begin{eqnarray*}
s & = & a x \: + \: b y \\
t & = & c x \: + \: d y \\
u & = & e \\
v & = & 0
\end{eqnarray*}
for $a$, $b$, $c$, $d$, $e$ complex numbers which we can be identified
with homogeneous coordinates on the linear sigma model moduli space.
The moduli space is determined by a pair of ${\bf C}^{\times}$
actions, whose actions on the homogeneous coordinates $a$, $b$, $c$, 
$d$, $e$ are determined by the actions on the homogeneous coordinates
$s$, $t$, $u$, $v$ on ${\bf F}_n$.
If we denote the ${\bf C}^{\times}$ actions on the moduli space
by $(\overline{\lambda}, \overline{\mu})$, then the homogeneous
coordinates $a$, $b$, $c$, $d$, $e$ have weights given by
\begin{center}
\begin{tabular}{c|ccccc}
 & $a$ & $b$ & $c$ & $d$ & $e$ \\ \hline
$\overline{\lambda}$ & $1$ & $1$ & $1$ & $1$ & $n$ \\
$\overline{\mu}$ & $0$ & $0$ & $0$ & $0$ & $1$ \\
\end{tabular}
\end{center}
The obstruction sheaf on the moduli space, computed by
the methods given earlier, is given by
\begin{displaymath}
H^1\left( {\bf P}^1, {\cal O}(-n) \right) \otimes_{ {\bf C} }
{\cal O}(0,1) \: = \: \bigoplus_1^{n-1} {\cal O}(0,1).
\end{displaymath}
But ${\cal O}(0,1)$ is the trivial bundle on ${\cal M}$ since the apparent
freedom of $e$ has been removed as noted earlier.  Thus the obstruction
bundle is ${\cal O}^{n-1}$.  Since the euler class of ${\cal O}^{n-1}$ is
trivial for $n\ge2$, all correlation functions will vanish.  This was to
be expected, as the virtual dimension is $E\cdot c_1({\bf F}_n)+2=4-n$,
so that all three~point functions vanish for $n\ge2$ by simple dimension
considerations.

\subsection{Bundles presented as kernels}

In this section, let us consider a bundle described in linear sigma
model language as a kernel.  We shall check that if it obeys the
usual physical constraints, then the corresponding sheaf on the moduli
space has the desired $c_1$.

\subsubsection{An example on ${\bf P}^{N-1}$}

Let us begin by studying examples of bundles on ${\bf P}^{N-1}$.
Consider a bundle ${\cal E}$ on ${\bf P}^{N-1}$ defined by
\begin{displaymath}
0 \: \longrightarrow \: {\cal E} \: \longrightarrow \:
\oplus_a {\cal O}(n_a) \: \stackrel{F^a_i}{ \longrightarrow} \:
\oplus_i {\cal O}(m_i) \: \longrightarrow \: 0
\end{displaymath}
where all the $n_a$ and $m_i$ are positive.

As before, the bundle ${\cal F}$ on the (linear sigma model) moduli
space of maps of degree $d$ is given as another kernel,
expressed between two bundles on the moduli space constructed
from the zero modes of the worldsheet fields corresponding to the
bundles $\oplus {\cal O}(n_a)$ and $\oplus {\cal O}(m_i)$.
The weights under the toric action of the zero modes are the same
as the weights of the original fields, hence
\begin{displaymath}
{\cal F} \: = \:
\mbox{ker } \left\{ \:
\oplus_a H^0\left( {\bf P}^1, {\cal O}(n_a d) \right) \otimes_{ {\bf C} }
{\cal O}(n_a) \: \longrightarrow \:
\oplus_i H^0\left( {\bf P}^1, {\cal O}(m_i d) \right) \otimes_{ {\bf C} }
{\cal O}(m_i) \: \right\}. 
\end{displaymath}
The maps are determined by the maps defining ${\cal E}$, in the
same fashion as for cokernels.  More specifically, we expand
out maps into homogeneous coordinates on ${\bf P}^{N-1}$ in terms
of homogeneous coordinates $u$, $v$ on the worldsheet,
then expand out the maps $F^a_i$ in terms of those $u$, $v$
in order to get the precise maps above.  We shall assume that the
resulting map is surjective, for simplicity.

First, let us check that ${\cal F}$, as we have defined it above,
has the correct rank.  On general principles, as outlined elsewhere,
the virtual rank of ${\cal F}$ is given by
\begin{displaymath}
c_1( \phi^* {\cal E} ) \: + \: (\mbox{rk }{\cal E})(1-g)
\end{displaymath}
and $g=0$ in the present case.
Our ${\cal F}$ above has rank
\begin{eqnarray*}
\lefteqn{\sum_a h^0\left( {\bf P}^1, {\cal O}(n_a d) \right) \: - \:
\sum_i h^0\left( {\bf P}^1, {\cal O}(m_i d) \right)} \\
& = & \sum_a (n_a d + 1) - \sum_i (m_i d + 1 ) \\
& = & d \left( \sum_a n_a \: - \: \sum_i m_i \right)
\: + \: \left( \sum_a 1 \: - \: \sum_i 1 \right).
\end{eqnarray*}
First, note that
\begin{displaymath}
d \left( \sum_a n_a \: - \: \sum_i m_i \right) \: = \: 
c_1(\phi^* {\cal E} )
\end{displaymath}
and also
\begin{displaymath}
\mbox{rank } {\cal E} \: = \: \sum_a 1 \: - \: \sum_i 1.
\end{displaymath}
Thus, plugging back into the expression above, we find that
the rank of ${\cal F}$ is the same as the predicted virtual rank of
${\cal F}$.

Next, we shall compute $c_1({\cal F})$, and demonstrate
that when ${\cal E}$ satisfies anomaly cancellation and has $c_1({\cal E})
= c_1(T {\bf P}^{N-1} )$, that $c_1({\cal F}) = c_1(T {\cal M})$,
as desired.

If we let $J$ denote the generator of the integral cohomology ring
of the (linear sigma model) moduli space 
${\cal M} = {\bf P}^{N(d+1) - 1}$, then we have that
\begin{eqnarray*}
c_1({\cal F}) & =  &
\sum_a \left( h^0({\bf P}^1, {\cal O}(n_a d) ) \right) n_a J \: - \:
\sum_i \left( h^0({\bf P}^1, {\cal O}(m_i d) ) \right) m_i J \\
& = & \sum_a n_a(n_a d + 1 ) J \: - \:
\sum_i m_i ( m_i d + 1 ) J \\
& = & \left( \sum_a n_a - \sum_i m_i \right) J \: + \:
\left( \sum_a n_a^2 - \sum_i m_i^2 \right) d J.
\end{eqnarray*}
Now, it is easy to compute that
\begin{displaymath}
c_1({\cal E}) \: = \: \left( \sum_a n_a \: - \: \sum_i m_i \right) J'
\end{displaymath}
(where $J'$ generates the cohomology ring of ${\bf P}^{N-1}$)
so $c_1({\cal E}) = c_1( T {\bf P}^{N-1} )$ implies that
\begin{displaymath}
\left( \sum_a n_a \: - \: \sum_i m_i \right) \: = \: N.
\end{displaymath}
Similarly, if we use $\mbox{ch}_2 = \frac12 c_1^2 -  c_2$,
then it is straightforward to compute that
\begin{displaymath}
\mbox{ch}_2( {\cal E} ) \: = \: \frac12 \left( \sum_a n_a^2 \: - \:
\sum_i m_i^2 \right) J'^2,
\end{displaymath}
so using the anomaly cancellation condition we find that
\begin{eqnarray*}
\sum_a n_a^2 \: - \: \sum_i m_i^2 & = &
N^2 \: - \: 2 \left( \begin{array}{c}
                     N \\ 2 \end{array} \right) \\
& = & N.
\end{eqnarray*}
Plugging these results back into the expression for $c_1( {\cal F})$,
we find that
\begin{eqnarray*}
c_1( {\cal F} ) & = & N J \: + \: N d J\\
& = & N(d+1) J \\
& = & c_1( T {\bf P}^{N(d+1)-1} = T {\cal M} ),
\end{eqnarray*}
exactly as desired.

\subsubsection{General analysis of bundles presented as kernels}

More generally, consider a bundle ${\cal E}$ over a toric variety $X$
defined as the kernel of the short exact sequence
\begin{displaymath}
0 \: \longrightarrow \: {\cal E} \: \longrightarrow
\: \oplus_a {\cal O}(\vec{n}_a )
\: \longrightarrow \: \oplus_r {\cal O}(\vec{m}_r ) \:
\longrightarrow \: 0.
\end{displaymath}
Over a linear sigma model moduli space of maps of degree $\vec{d}$,
the short exact sequence above induces the long exact sequence
\begin{displaymath}
\begin{array}{ccccccc}
0 & \longrightarrow & {\cal F} & \longrightarrow &
\oplus_a H^0\left( {\bf P}^1, {\cal O}(\vec{n}_a \cdot \vec{d} )
\right) \otimes_{ {\bf C} } {\cal O}(\vec{n}_a) & \longrightarrow &
\oplus_r H^0\left( {\bf P}^1, {\cal O}(\vec{m}_r \cdot \vec{d} ) \right)
\otimes_{ {\bf C} } {\cal O}(\vec{m}_r) \\
 & \longrightarrow & {\cal F}_1 & \longrightarrow &
\oplus_a H^1\left( {\bf P}^1, {\cal O}(\vec{n}_a \cdot \vec{d} )
\right) \otimes_{ {\bf C} } {\cal O}(\vec{n}_a) & \longrightarrow &
\oplus_r H^1\left( {\bf P}^1, {\cal O}(\vec{m}_r \cdot \vec{d} ) \right)
\otimes_{ {\bf C} } {\cal O}(\vec{m}_r) \\
 & \longrightarrow & 0. & & & &  
\end{array}
\end{displaymath}
It is straightforward to check, using the same methods as
for bundles presented as cokernels, that on the open subset
of ${\cal M}$ corresponding to honest maps,
${\cal F}$ and ${\cal F}_1$ match $R^0 \pi_* \alpha^* {\cal E}$
and $R^1 \pi_* \alpha^* {\cal E}$, respectively.

Let us check that the sheaves ${\cal F}$, ${\cal F}_1$ have the expected
total rank predicted in section~\ref{excesszeromodes}.
Note that
\begin{eqnarray*}
\mbox{rank } \left( {\cal F} \ominus {\cal F}_1 \right) & = &
\sum_{ \vec{n}_a \cdot \vec{d} \geq 0 } \left( \vec{n}_a \cdot \vec{d} \: + \:
1 \right) \: - \:
\sum_{ \vec{m}_r \cdot \vec{d} \geq 0 } \left( \vec{m}_r \cdot \vec{d} 
\: + \: 1 \right) \\
& & \: - \: \sum_{ \vec{n}_a \cdot \vec{d} < 0 } \left(
- \vec{n}_a \cdot \vec{d} \: - \: 1 \right) \: + \:
\sum_{ \vec{m}_r \cdot \vec{d} < 0 } \left(
- \vec{m}_r \cdot \vec{d} \: - \: 1 \right) \\
& = & \left( \sum_a 1 \: - \: \sum_r 1 \right) \: + \:
\left( \sum_a \vec{n}_a \cdot \vec{d} \: - \:
\sum_r \vec{m}_r \cdot \vec{d} \right) \\
& = & \mbox{rank } {\cal E} \: + \: c_1\left( \phi^* {\cal E} \right) \\
& = & \chi\left( \phi^* {\cal E} \right)
\end{eqnarray*}
precisely as desired.

Next, let us check that the condition
\begin{displaymath}
\Lambda^{top} {\cal F}^{\vee} \otimes \Lambda^{top} {\cal F}_1 \: 
\cong \: K_{ {\cal M} } \otimes \Lambda^{top} {\cal G}_1,
\end{displaymath}
as derived formally in section~\ref{excesszeromodes},
is indeed satisfied by the sheaves ${\cal F}$, ${\cal F}_1$ above.
Present the tangent bundle to the toric variety  in the form
\begin{displaymath}
0 \: \longrightarrow \: {\cal O}^{\oplus m} \: \longrightarrow \:
\oplus_n {\cal O}(\vec{q}_n ) \: \longrightarrow \: TX \:
\longrightarrow \: 0.
\end{displaymath}
The anomaly cancellation condition in the (0,2) gauged linear sigma
model has the form
\begin{displaymath}
\sum_a n_a^i n_a^j \: - \: \sum_r m_r^i m_r^j \: = \: \sum_n q_n^i q_n^j
\end{displaymath}
for all $i$, $j$, and the constraint that $\Lambda^{top} {\cal E}^{\vee}
\cong K_X$ implies that
\begin{displaymath}
\sum_a n_a^i \: - \: \sum_r m_r^i  \: = \:
\sum_n q_n^i 
\end{displaymath}
for each $i$.  
Then, from the long exact sequence above, we see that
\begin{eqnarray*}
c_1\left( {\cal F} \ominus {\cal F}_1 \right) & = &
\sum_{ \vec{n}_a \cdot \vec{d} \geq 0 } \left( \vec{n}_a \cdot \vec{d} \: + \:
1 \right) n_a^i J_i  \: - \: \sum_{ \vec{m}_r \cdot \vec{d} \geq 0 }
\left( \vec{m}_r \cdot \vec{d} \: + \: 1 \right) m_r^i J_i \\
& & \: - \: \sum_{\vec{n}_a \cdot \vec{d} < 0 }\left( - \vec{n}_a \cdot \vec{d}
\: - \: 1 \right) n_a^i J_i 
\: + \: \sum_{ \vec{m}_r \cdot \vec{d} < 0 } \left( - \vec{m}_r \cdot \vec{d}
\: - \: 1 \right) m_r^i J_i \\
& = & \left( \sum_a n_a^i J_i \right) \: + \:
\left( \sum_a \left( \vec{n}_a \cdot \vec{d} \right) n_a^i J_i \right) \: - \:
\left( \sum_r m_r^i J_i \right) \: - \:
\left( \sum_r \left( \vec{m}_r \cdot \vec{d} \right) m_r^i J_i \right).
\end{eqnarray*}
For the tangent bundle, we similarly obtain
\begin{displaymath}
c_1\left( T {\cal M} \ominus {\cal G}_1 \right) \: = \:
\left( \sum_n q_n^i J_i \right) \: + \:
\left( \sum_n \left( \vec{q}_n \cdot \vec{d} \right) q_n^i J_i \right).
\end{displaymath}
Thus, using the anomaly cancellation condition and the statement that
$\Lambda^{top} {\cal E}^{\vee} \cong K_X$, we see that
\begin{displaymath}
c_1\left( {\cal F} \ominus {\cal F}_1 \right) \: = \:
c_1\left( T {\cal M} \ominus {\cal G}_1 \right),
\end{displaymath}
precisely as predicted on general grounds in section~\ref{excesszeromodes}.

\subsection{Bundles presented as the cohomology of a monad}

The previous discussion also extends to bundles presented as monads,
as we shall now briefly outline.
In this case, the monad is a short complex of the form
\begin{displaymath}
0 \: \longrightarrow \: {\cal O}^{\oplus k} \:
\stackrel{E}{\longrightarrow } \: \oplus_a {\cal O}(\vec{n}_a) \:
\stackrel{F}{\longrightarrow} \: \oplus_i {\cal O}(\vec{m}_i) \:
\longrightarrow \: 0
\end{displaymath}
and the bundle ${\cal E}$ is defined by
\begin{displaymath}
{\cal E} \: \equiv \: \frac{ \mbox{ker } F }{\mbox{im }E}.
\end{displaymath}
Physically, this is realized via (0,2) fermi multiplets
$\Lambda^a$ of charge $\vec{n}_a$, (0,2) chiral multiplets
$p_i$ of charge $\vec{m}_i$, and neutral (0,2) chiral multiplets
$\Sigma^{\lambda}$, together with a superpotential term realizing
$F$, and susy transformations realizing $E$.

For simplicity, we break the cohomology of the complex
into a pair of short exact sequences
\begin{eqnarray}
0 \: \longrightarrow \: \mbox{ker } F \:
\longrightarrow & \oplus_a {\cal O}(\vec{n}_a) & 
\stackrel{F}{\longrightarrow} \: \oplus_i {\cal O}(\vec{m}_i) \:
\longrightarrow \: 0,  \label{eq1} \\
0 \: \longrightarrow \: {\cal O}^{\oplus k} \: 
\stackrel{E}{\longrightarrow} & \mbox{ker }F & \:
\longrightarrow \: {\cal E} \: \longrightarrow \: 0.
\label{eq2}
\end{eqnarray}

Expanding fields into their zero modes, we see that
the short exact sequence~(\ref{eq1}) induces
\begin{displaymath}
\begin{array}{ccccccc}
0 & \longrightarrow &
\widetilde{ \left( \mbox{ker }F \right)}_0 & \longrightarrow &
\oplus_a H^0\left( {\bf P}^1, {\cal O}(\vec{n}_a \cdot \vec{d} ) \right)
\otimes_{ {\bf C} } {\cal O}(\vec{n}_a) &
\longrightarrow &
\oplus_i H^0\left({\bf P}^1, {\cal O}(\vec{m}_i \cdot \vec{d}) \right)
\otimes_{ {\bf C} } {\cal O}(\vec{m}_i) \\
 & \longrightarrow &
\widetilde{ \left( \mbox{ker }F \right)}_1 &
\longrightarrow &
\oplus_a H^1\left( {\bf P}^1, {\cal O}(\vec{n}_a \cdot \vec{d} ) \right)
\otimes_{ {\bf C} } {\cal O}(\vec{n}_a) & \longrightarrow &
\oplus_i H^1\left({\bf P}^1, {\cal O}(\vec{m}_i \cdot \vec{d}) \right)
\otimes_{ {\bf C} } {\cal O}(\vec{m}_i) \\
& \longrightarrow & 0 & & & & 
\end{array}
\end{displaymath}
and the short exact sequence~(\ref{eq2}) induces
\begin{displaymath}
0 \: \longrightarrow \: {\cal O}^{\oplus k} \:
\longrightarrow \: 
\widetilde{ \left( \mbox{ker }F \right)}_0
\: \longrightarrow \: {\cal F} \: \longrightarrow 0, 
\end{displaymath}
\begin{displaymath}
{\cal F}_1 \: \cong \: \widetilde{ \left( \mbox{ker }F \right)}_1.
\end{displaymath}

It is now straightforward to compute that
\begin{eqnarray*}
\mbox{rank } {\cal F} \: - \: \mbox{rank } {\cal F}_1 & = &
\mbox{rank } {\cal E} \: + \: c_1\left( \phi^* {\cal E} \right), \\
c_1({\cal F}) \: - \: c_1\left( {\cal F}_1 \right) & = &
c_1\left( T {\cal M} \right) \: - \:
c_1 \left( {\cal G}_1 \right)
\end{eqnarray*}
exactly as desired.

\subsection{A technical aside}   \label{multpres}

A linear sigma model describes a physically natural compactification
of moduli spaces of maps, and $(0,2)$ linear sigma models describe
physically natural extensions of the sheaves $R^{i} \pi_* \alpha^*
{\cal E}$ over
the compactifications of the moduli spaces.  However,
we shall see in this section that
the precise extensions depend upon the presentation of the
gauge bundle ${\cal E}$, and not all presentations yield
consistent linear sigma models.

In particular, it is important for our construction that the
linear sigma model anomaly cancellation condition be met,
which is somewhat stronger than the mathematical
statement $\mbox{ch}_2({\cal E}) = \mbox{ch}_2(TX)$.
We shall see that the linear sigma model
anomaly cancellation condition can even distinguish different
presentations of the same gauge bundle.
Presentations that fail the linear sigma model anomaly cancellation
condition have different extensions over the compactification
divisor which do not have desirable properties.

If we schematically let charges of left-moving fermions be denoted
$\vec{n}_a$ and charges of right-moving fermions be denoted $\vec{q}_s$,
then we have imposed two conditions, namely
\begin{eqnarray*}
\sum_a n_a^i & = & \sum_s q_s^i \: \mbox{ for each } i, \\
\sum_a n_a^i n_a^j & = & \sum_s q_s^i q_s^j \: \mbox{ for each }i, j
\end{eqnarray*}
The first of these statements is the linear-sigma-model version
of the requirement $\Lambda^{top} {\cal E}^{\vee} \cong K_X$,
and the second \cite{dgm} is the linear-sigma-model version of the 
anomaly cancellation condition $\mbox{ch}_2({\cal E}) = \mbox{ch}_2(TX)$.

Our linear-sigma-model requirements are slightly stronger than
their mathematical counterparts.  For example, consider the tangent
bundle of ${\bf P}^1$, presented in its physically canonical form as
\begin{displaymath}
0 \: \longrightarrow \: {\cal O} \: \longrightarrow \:
{\cal O}(1)^2 \: \longrightarrow \: T {\bf P}^1 \: \longrightarrow \: 0
\end{displaymath}
Suppose the gauge bundle ${\cal E}$ is described by a single free
fermion of charge 2.  Mathematically, this is the statement that
${\cal E} = {\cal O}(2)$, which is isomorphic to
$T {\bf P}^1$.  Since ${\cal E} \cong T {\bf P}^1$,
we trivially have the statement that $\mbox{ch}_2({\cal E}) = 
\mbox{ch}_2(T {\bf P}^1)$.  However, this setup no longer satisfies
the linear sigma model anomaly cancellation condition:
\begin{eqnarray*}
\sum_s \left( q_s \right) \left( q_s \right) & = & \sum_1^2
(1)^2 \: = \: 2 \\
\sum_a \left( n_a \right) \left( n_a \right) & = & \sum_1^1
(2)^2 \: = \: 4
\end{eqnarray*}
and so physically, the bundle ${\cal E}$ described by a single
free fermion of charge 2 on ${\bf P}^1$ does not describe
a consistent linear sigma model, even though mathematically
$\mbox{ch}_2 ( {\cal E} ) = \mbox{ch}_2 (TX)$.

Let us apply our construction to this inconsistent linear sigma model.
If we present the tangent bundle of ${\bf P}^1$ in
its physically canonical form, as a pair of 
left-moving fermions of charge 1 with a `gauged fermionic symmetry,'
the resulting sheaf ${\cal F}$ over
the moduli space is the tangent bundle to the moduli space.
However, if we present it as merely ${\cal O}(2)$,
{\it i.e.} as a single free fermion of charge 2,
which physically is not the canonical presentation, then the resulting
sheaf ${\cal F}$ over the moduli space is given by
\begin{displaymath}
H^0\left( {\bf P}^1 , {\cal O}(2d) \right) \otimes_{ {\bf C} } {\cal O}(2)
\: = \: \oplus_1^{2d+1} {\cal O}(2)
\end{displaymath}
which clearly is not only no longer
the tangent bundle
of the moduli space ${\bf P}^{2d+1}$,
but in fact does not even have the property
$\Lambda^{top} {\cal F}^{\vee} \cong K_{ {\cal M} }$.
This ${\cal F}$ is a very different extension of $R^0 \pi_* \alpha^* {\cal E}$.

Thus, we see in this example that the gauged linear sigma model
anomaly cancellation condition, which is slightly stronger than
merely $\mbox{ch}_2({\cal E}) = \mbox{ch}_2(TX)$, plays an
important role in the consistency of our construction.
When we applied our construction to an inconsistent linear sigma model,
the induced sheaves ${\cal F}$, ${\cal F}_1$ do not have correct Chern
classes.  Furthermore, this also shows that the extensions across the
compactification divisor of
$R^{i} \pi_* \alpha^* {\cal E}$ defined by our construction
depend upon the presentation of ${\cal E}$.

The reader might object that this phenomenon is too degenerate,
in that ${\bf P}^1$ only has dimension one,
and so can not provide good examples for subtleties revolving
around second Chern classes.  However, there are more examples in
higher dimensions.

The tangent bundle of ${\bf P}^1 \times {\bf P}^1$ gives a slightly
less degenerate
example of this phenomenon.  There are three presentations of this
bundle:
\begin{enumerate}
\item The standard (2,2) presentation, as a cokernel
\begin{displaymath}
0 \: \longrightarrow \: {\cal O}^{\oplus 2} \:
\longrightarrow \:
{\cal O}(1,0)^2 \oplus {\cal O}(0,1)^2 \: \longrightarrow \:
T \left( {\bf P}^1 \times {\bf P}^1 \right) \: \longrightarrow \: 0
\end{displaymath}
corresponding physically to four charged fermi superfields 
(part of the (2,2) chiral superfields describing the homogeneous
coordinates) and two
neutral chiral superfields (part of the (2,2) gauge superfields
corresponding to the two $U(1)$'s).
\item A cokernel presentation for one $T {\bf P}^1$,
and a free fermion presentation for the other:
\begin{displaymath}
0 \: \longrightarrow \:
{\cal O} \: \longrightarrow \:
{\cal O}(1,0)^2 \oplus {\cal O}(0,2) \: \longrightarrow \: 
T \left( {\bf P}^1 \times {\bf P}^1 \right) \: \longrightarrow \: 0.
\end{displaymath}
\item A free fermion presentation for both sides:
\begin{displaymath}
T \left( {\bf P}^1 \times {\bf P}^1 \right) \: \cong \:
{\cal O}(2,0) \oplus {\cal O}(0,2).
\end{displaymath}
\end{enumerate}
Only the standard (2,2) presentation satisfies the strong
linear sigma model anomaly cancellation condition for a gauge
bundle on ${\bf P}^1 \times {\bf P}^1$; the second two presentations
do not satisfy this condition.
As a result, each of these physical presentations of the tangent bundle
gives a different induced sheaf ${\cal F}$ over the compactified
moduli space, and only for the $(2,2)$ presentation does the
resulting induced sheaf have desirable properties.

The first presentation yields ${\cal F}$ isomorphic to the tangent
bundle of the moduli space ${\bf P}^{2d_1 +1} \times {\bf P}^{2 d_2 + 1}$.

The second presentation yields
\begin{displaymath}
0 \: \longrightarrow \:
{\cal O} \: \longrightarrow \:
\oplus_1^2 \left( H^0\left( {\bf P}^1, {\cal O}(d_1) \right) \otimes_{ {\bf C} }
{\cal O}(1,0) \right) \oplus \left( 
H^0\left( {\bf P}^1, {\cal O}(2d_2) \right) \otimes_{ {\bf C} }
{\cal O}(0,2) \right)
\: \longrightarrow \: {\cal F} \: \longrightarrow \: 0
\end{displaymath}
or, equivalently,
\begin{displaymath}
{\cal F} \: \cong \:
\pi_1^* T {\bf P}^{2d_1+1} \oplus \oplus_1^{2d_2+1} {\cal O}(0,2).
\end{displaymath}

The third presentation yields
\begin{eqnarray*}
{\cal F} & = & \left(
H^0\left( {\bf P}^1 , {\cal O}(2d_1) \right) \otimes_{ {\bf C} }
{\cal O}(2,0) \right)
\oplus
\left( H^0\left( {\bf P}^1, {\cal O}(2d_2) \right) \otimes_{ {\bf C} }
{\cal O}(0,2) \right) \\
& = & \oplus_1^{2d_1+1} {\cal O}(2,0) \oplus
\oplus_1^{2d_2+1} {\cal O}(0,2).
\end{eqnarray*}

So long as $d_1$ and $d_2$ are both positive, ${\cal F}_1 \equiv 0$ in
each case.

Note that for each presentation, 
\begin{eqnarray*}
\mbox{rank } {\cal F} & = & \left( 2d_1 \: + \: 1 \right)
\: + \: \left( 2 d_2 \: + \: 1 \right) \\
& = & \mbox{rank } T \left( {\bf P}^1 \times {\bf P}^1 \right) \: + \:
c_1 \left( \phi^* T \left( {\bf P}^1 \times {\bf P}^1 \right) \right).
\end{eqnarray*}

However, although the ranks match, it is straightforward to check
that, just as for $T {\bf P}^1$, the first Chern classes do not match,
and (except for the first presentation) do not have the desired symmetry
property $\Lambda^{top} {\cal F}^{\vee} \cong K_{ {\cal M} }$.
The reason for this is the same as for ${\bf P}^1$: 
the linear sigma model anomaly cancellation condition is only met
for the first presentation.

Thus, we see in these examples several important technicalities of
linear sigma models and our constructions:
\begin{itemize}
\item First, different presentations of a single, fixed gauge bundle ${\cal E}$
can define different extensions of $R^{i} \pi_* \alpha^* {\cal E}$
across the compactified moduli space.
\item Second, the linear sigma model anomaly cancellation condition
can distinguish different presentations of the same bundle.
\item Third, if a given presentation of the gauge bundle fails
the linear sigma model anomaly cancellation condition, then
even if there exists an alternate
presentation that satisfies that condition, the induced sheaves ${\cal F}$,
${\cal F}_1$
determined by the failing presentation have the wrong Chern classes
to be useful in our calculations of correlation functions.
\end{itemize}

\section{Verification of results of Adams-Basu-Sethi}  \label{checkabs}

In this section we shall give first-principles verifications of
some conjectures made in \cite{abs}.

\subsection{Physical prediction for quantum cohomology}

One of the first examples of a heterotic quantum cohomology ring\footnote{
As emphasized earlier, in a heterotic theory, one will not be able
to make sense of a ring structure in general,
but in special cases -- such as the deformation of the tangent bundle
considered here -- such a structure can still be meaningful.}
considered in \cite{abs}[section 6.2] concerns a heterotic theory on
${\bf P}^1 \times {\bf P}^1$, with gauge bundle given by a deformation
of the tangent bundle.
Recall that we can describe the tangent bundle of ${\bf P}^1 \times
{\bf P}^1$ as the cokernel of the map
\begin{displaymath}
{\cal O} \oplus {\cal O} \stackrel{ {\scriptsize
\left[ \begin{array}{cc}
       x_1 & 0 \\
       x_2 & 0 \\
        0 & \widetilde{x_1} \\
        0 & \widetilde{x_2} \end{array} \right] }}{\longrightarrow}
{\cal O}(1,0)^2 \oplus {\cal O}(0,1)^2
\end{displaymath}
where $x_1$, $x_2$ are homogeneous coordinates on the first ${\bf P}^1$,
and $\widetilde{x_1}$, $\widetilde{x_2}$ are homogeneous coordinates
on the second ${\bf P}^1$.
In \cite{abs}[section 6.2] a deformation of the tangent bundle of
${\bf P}^1 \times {\bf P}^1$ is described as the cokernel of the map
\begin{displaymath}
{\cal O} \oplus {\cal O} \stackrel{ {\scriptsize
\left[ \begin{array}{cc}
       x_1 & \alpha_1 x_1 + \alpha_2 x_2 \\
       x_2 & \alpha'_1 x_1 + \alpha'_2 x_2 \\
        \beta_1 \widetilde{x_1} + \beta_2 \widetilde{x_2} & \widetilde{x_1} \\
        \beta'_1 \widetilde{x_1} + \beta'_2 \widetilde{x_2} & \widetilde{x_2} \end{array} \right] }}{\longrightarrow}
{\cal O}(1,0)^2 \oplus {\cal O}(0,1)^2
\end{displaymath}
where $\alpha_i$, $\alpha'_i$, $\beta_i$, $\beta'_i$ are constants.

In the special case that $\alpha_1 = \epsilon_1$ and $\alpha'_2 = \epsilon_2$,
and all other constants vanish,
the authors of \cite{abs} obtain what can be described as a deformation of the
chiral ring relations for a $(2,2)$ model on ${\bf P}^1 \times {\bf P}^1$,
namely
\begin{eqnarray*}
\tilde{X}^2 & = & \exp(i t_2 ) \\
X^2 \: - \: \left( \epsilon_1 - \epsilon_2 \right) X \tilde{X}  & = &
\exp(i t_1)
\end{eqnarray*}
where $t_1$, $t_2$ are K\"ahler parameters describing the sizes of the
${\bf P}^1$'s, and $X$, $\tilde{X}$ are degree two generators.
(The sign ambiguity is meaningless, because $\epsilon_1$, $\epsilon_2$
act as homogeneous coordinates on the deformation space.)
Note in the special case of the $(2,2)$ locus, where $\epsilon_1=\epsilon_2$,
the relations above reduce precisely to the ordinary
quantum chiral ring relations for ${\bf P}^1 \times {\bf P}^1$. 

In terms of correlation functions, the quantum cohomology claim
above allows us to compute quantum corrections in terms of classical
results.  We shall describe the classical computation in section~\ref{compute}.
Without even doing the classical computation,
it is not hard to see that consistency\footnote{
Write $<X^2> = g$ for some function $g$ of $\epsilon_1$, $\epsilon_2$,
with the other classical correlation functions as listed.
One can then compute $<X \tilde{X}^3 > = \exp(it_2)$
and $<X^2 \tilde{X}^2> = g \exp(it_2)$.  Then, plug those values
into $<\tilde{X}^2\left( X^2 - (\epsilon_1 - \epsilon_2) X \tilde{X}
\right)> = 0$
to find that $g = \epsilon_1 - \epsilon_2$.
} of the ring structure
forces a nonzero value for $\langle X^2\rangle$ when $\epsilon_1\ne
\epsilon_2$ (i.e.\ off the $(2,2)$ locus) while the other classical
correlation functions are unchanged.  In any event, the
classical correlation functions can be shown to be given by 
\begin{equation}
\label{classcorr}
< \tilde{X}^2 > \: = \: <1> \: = \: 0, \: \: \:
< X \tilde{X} > \: = \: 1, \: \: \: < X^2 > \: = \: \epsilon_1-\epsilon_2.
\end{equation}
We can formally compute a higher-order correlator by using the quantum
cohomology relations, {\it e.g.}
\begin{eqnarray}
< X \tilde{X}^3 > & = & < \left( X \tilde{X} \right) \tilde{X}^2 > \\
& = & 
< X \tilde{X} > \exp(it_2) \: = \: \exp(it_2) \label{xyyy} \\
< X \tilde{X}^5 > & = & <X \tilde{X} \left( \tilde{X}^2
\right)^2 > \: = \: \exp(2it_2) \\
< X \tilde{X}^7 > & = & \exp(3it_2) \\
< \tilde{X}^4 > & = & <1> \exp(2it_2) \: = \: 0 \label{yyyy} \\
< X^2 \tilde{X}^2 > & = & <X^2> \exp(it_2) \: = \: 
 (\epsilon_1-\epsilon_2) \exp(it_2) \label{xxyy} \\
< X \tilde{X} \left( X^2 - (\epsilon_1 - \epsilon_2) X \tilde{X} \right)
> & = & < X \tilde{X} > \exp(i t_1) \: = \: \exp(i t_1) \\
< X^3 \tilde{X} > & = & \exp(it_1) + (\epsilon_1-\epsilon_2)^2
 \exp(it_2)\label{xxxy} \\
< X^4 > & = & < X^2 \left( \exp(it_1) \:+  \: \left( \epsilon_1 -
\epsilon_2 \right) X \tilde{X} \right) > \\
& = &
2 \left( \epsilon_1 - \epsilon_2 \right) \exp(it_1) 
 + (\epsilon_1-\epsilon_2)^3 \exp(it_2)\label{xxxx}
\end{eqnarray}
and so forth.

In section~\ref{compute}, we shall directly compute the four-point
functions listed above, and we will directly confirm that
they do have the form listed.
In general, one would only expect to
obtain the answer up to a coordinate change, of course,
but in this simple example, we shall recover the advertised
four-point functions.

One quick physical way to derive the quantum cohomology statement above 
(following \cite{daveronen},
but slightly different from \cite{abs}) is to compute the
one-loop effective superpotential
\begin{eqnarray*}
\tilde{W}_{eff} & = & \Upsilon_1\left( i \hat{\tau}_1 \: - \:
\frac{1}{2 \pi} \log\left( \frac{ \sigma_1 \: + \: \epsilon_1 \sigma_2}{
\Lambda} \right) \: - \:
\frac{1}{2\pi}\log\left( \frac{ \sigma_1 \: + \: \epsilon_2 \sigma_2 }{
\Lambda} \right) \right) \\
& & \: + \: \Upsilon_2 \left( i \hat{\tau}_2 \: - \:
\frac{1}{2\pi} \log \left( \frac{\sigma_2}{\Lambda} \right) \: - \:
\frac{1}{2 \pi} \log \left( \frac{\sigma_2}{\Lambda} \right) \right)
\end{eqnarray*}
(where the $\Upsilon$'s are (0,2) gauge multiplets),
from which setting 
\begin{displaymath}
\frac{ \partial \tilde{W}_{eff} }{\partial \Upsilon_a } \: = \: 0
\end{displaymath}
gives the relations
\begin{eqnarray*}
\left( \sigma_1 \: + \: \epsilon_1 \sigma_2 \right)
\left( \sigma_1 \: + \: \epsilon_2 \sigma_2 \right)
& = & q_1 \\
\sigma_2^2 & = & q_2
\end{eqnarray*}
which, after a change of variables, are equivalent to the relations
given in \cite{abs}.

In the case of interest, the gauge bundle ${\cal E}$ is given by
\begin{equation}
\label{eq:e}
0 \: \longrightarrow \: 
{\cal O} \oplus {\cal O} \stackrel{ {\scriptsize
\left[ \begin{array}{cc}
       x_1 & \epsilon_1 x_1 \\
       x_2 & \epsilon_2 x_2 \\
        0 & \widetilde{x_1} \\
        0 & \widetilde{x_2} \end{array} \right] }}{\longrightarrow} \:
{\cal O}(1,0)^2 \oplus {\cal O}(0,1)^2
\: \longrightarrow \: {\cal E} \: \longrightarrow \: 0.
\end{equation}

For degree $(1,0)$ maps, the linear sigma model moduli space is
${\bf P}^3 \times {\bf P}^1$.  If we let $\alpha_0$, $\alpha_1$,
$\alpha'_0$, $\alpha'_1$ denote homogeneous coordinates on the
${\bf P}^3$ (obtained from expanding out the homogeneous coordinates
$x_1$, $x_2$ in terms of their zero modes), and let $\beta_0$, $\beta_1$
denote the homogeneous coordinates on the ${\bf P}^1$, then the
sheaf ${\cal F}$ on the moduli space is described by
\begin{equation}
\label{eq:f}
0 \: \longrightarrow \:
{\cal O} \oplus {\cal O} \stackrel{ {\scriptsize
\left[ \begin{array}{cc}
       \alpha_0 & \epsilon_1 \alpha_0 \\
       \alpha_1 & \epsilon_1 \alpha_1 \\
       \alpha'_0 & \epsilon_2 \alpha'_0 \\
       \alpha'_1 & \epsilon_2 \alpha'_1 \\
       0 & \beta_0 \\
       0 & \beta_1 \end{array} \right] }}{\longrightarrow} \:
{\cal O}(1,0)^4 \oplus {\cal O}(0,1)^2 
\: \longrightarrow \: {\cal F} \: \longrightarrow \: 0.
\end{equation}

Next, we shall find polynomial representatives for the 
relevant sheaf cohomology groups.
{}From (\ref{eq:e}), we derive the long
exact sequence
\begin{equation}
\label{les}
\begin{array}{ccccccc}
0 & \longrightarrow & H^0\left(X, {\cal E}^{\vee} \right) &
\longrightarrow & H^0\left( {\cal O}(-1,0)^2 \oplus {\cal O}(0,-1)^2 \right)
& \longrightarrow & H^0\left( {\cal O}^2 \right) \\
 & \longrightarrow & H^1\left(X, {\cal E}^{\vee} \right)
& \longrightarrow &
H^1\left(  {\cal O}(-1,0)^2 \oplus {\cal O}(0,-1)^2 \right)
& \longrightarrow & \cdots
\end{array}
\end{equation}
from which we read off that
\begin{equation}
\label{basis}
H^1\left(X, {\cal E}^{\vee} \right) \: = \:
H^0\left( X, {\cal O}^2 \right) \: = \: {\bf C}^2.
\end{equation}

In particular, the two elements of $H^1(X, {\cal E}^{\vee})$ can both
be represented by constants.
Essentially the same calculation using (\ref{eq:f}) in place of (\ref{eq:e})
reveals that
\begin{displaymath}
H^1\left({\cal M}, {\cal F}^{\vee} \right) \: = \:
H^0\left( {\cal M}, {\cal O}^2 \right) \: = \: {\bf C}^2.
\end{displaymath}
We therefore have a natural map
\[
H^1(X,{\cal E}^\vee)\simeq {\bf C}^2\simeq H^1({\cal M},{\cal F}^\vee)
\]
which is the linear sigma model version of the maps $\psi_i$ 
(\ref{eval1}).

\subsection{Computation of the correlation functions} \label{compute}

Let us now explicitly compute the classical correlation functions
listed in equation~(\ref{classcorr}) and the four-point functions listed
in equations (\ref{xxxx}), (\ref{xxxy}), (\ref{xxyy}), (\ref{xyyy}),
and (\ref{yyyy}).  The classical contributions to these four-point functions
all vanish, and the only worldsheet instanton contributions can come
from the $(1,0)$ and $(0,1)$ sectors.  

Before we begin, we need to observe that the natural basis (\ref{basis}) 
for $H^2({\bf P}^1\times{\bf P}^1,{\cal E}^\vee)$ does not specialize to
the usual basis for $H^2({\bf P}^1\times{\bf P}^1)$ generated by the
hyperplane classes of the ${\bf P}^1$ factors on the $(2,2)$ locus.  
Let's denote the
basis determined by (\ref{basis}) as $Y,\tilde{Y}$. The 
$4\times 2$ matrix in (\ref{eq:e}), specialized to the $(2,2)$ locus
$\epsilon_1=\epsilon_2$, shows that $Y,\tilde{Y}$ is related to the 
standard quantum cohomology basis $X,\tilde{X}$ by
\begin{equation}
\label{substitution}
X = Y+\epsilon\tilde{Y}, \qquad
\tilde{X} = \tilde{Y},
\end{equation}
as can be inferred from the first two rows (resp.\ the last 2 rows).
Here we have put $\epsilon=\epsilon_1=\epsilon_2$.

To explain the method, we compute the classical correlation functions of
$Y$ and $\tilde{Y}$ in detail.  We start with the dual of (\ref{eq:e})
\begin{equation}
  \label{edual}
0\to {\cal E}^\vee\to {\cal O}(-1,0)^2\oplus{\cal O}(0,-1)^2
\stackrel{{\scriptsize
\left[ \begin{array}{cccc}
x_1 & x_2 & 0 & 0 \\
\epsilon_1 x_1 & \epsilon_2 x_2 & \tilde{x}_1 & \tilde{x}_2
\end{array}\right] }
}{\longrightarrow}
{\cal O}^2\to 0.
\end{equation}

Let us denote by $e_1,e_2,f_1,f_2$ the natural basis for ${\cal O}^4$, 
corresponding to the columns of the matrix in (\ref{edual}).
We want to compute the cohomology classes $Y,\tilde{Y}\in H^1({\cal
E}^\vee)$.  We do this by computing images of $(1,0)\ (0,1)\in H^0({\cal O}^2)=
{\bf C}^2$ of the coboundary mapping in (\ref{les}). 
We compute using the Cech cover $U_{ij}=\{x_i\ne0,\ \tilde{x}_j\ne0\}$ of
${\bf P}^1\times {\bf P}^1$, where $1\le i, j\le2$.

Let us first compute $\tilde{Y}$ as the coboundary of $(0,1)$.  First
we must lift $(0,1)$ to a section of ${\cal O}(-1,0)^2\oplus{\cal
O}(0,-1)^2$ over $U_{ij}$, and the simplest lift is $\tilde{x}_j^{-1} f_j$.
Then $\tilde{Y}$ has a Cech representative
\begin{equation}
  \label{cechyt}
\tilde{Y}_{ij,i'j'}=\tilde{x}_{j'}^{-1} f_{j'}-\tilde{x}_j^{-1} f_j.
\end{equation}
To make sense of (\ref{cechyt}) we fix an ordering of the sets in the open
cover, say $U_{11},\ U_{12},\ U_{21},\ U_{22}$.

Similarly, we lift $(1,0)$ to $x_i^{-1}e_i-\epsilon_i\tilde{x}_j^{-1}f_j$
on $U_{ij}$, yielding the Cech representative
\begin{equation}
  \label{cechy}
Y_{ij,i'j'}=x_{i'}^{-1}e_{i'}-x_i^{-1}e_i+\epsilon_i\tilde{x}_j^{-1}f_j-
\epsilon_{i'}\tilde{x}_{j'}^{-1}f_{j'}.
\end{equation}

To compute the classical correlator $\langle \tilde{Y}^2\rangle$, we
take the cup and wedge product of $\tilde{Y}$ with itself to get a
Cech representative of $H^2(\Lambda^2{\cal E}^\vee)$. From our
constraint (\ref{u1cond}) this is identified with 
an element of $H^2(K)\simeq {\bf
C}$ so we will get a number.  In the present situation of ${\bf P}^1\times
{\bf P}^1$, we have $K\simeq{\cal O}(-2,-2)$.

To compute explicitly, note the inclusion
\begin{equation}
  \label{wedge2in}
\Lambda^2{\cal E}^\vee\simeq {\cal O}(-2,-2)\hookrightarrow
\Lambda^2\left({\cal O}(-1,0)^2\oplus{\cal O}(0,-1)^2\right)\simeq
{\cal O}(-2,0)\oplus{\cal O}(-1,-1)^4\oplus{\cal O}(0,-2).
\end{equation}
Our strategy is to first compute the cup product as a representative of
$H^2(\Lambda^2({\cal O}(-1,0)^2\oplus{\cal O}(0,-1)^2)=
H^2({\cal O}(-2,0)\oplus{\cal O}(-1,-1)^4\oplus{\cal O}(0,-2))$ using
the inclusion (\ref{wedge2in}). Then we interpret
this cocycle as a representative of $H^2(
\Lambda^2{\cal E}^\vee)=H^2({\cal O}(-2,-2))$.  We do this explicitly
by computing the inclusion (\ref{wedge2in}).  So the computation proceeds
from (\ref{cechy}) and (\ref{cechyt}) by the explicit computation of
(\ref{wedge2in}) and the explicit computation of the cup product.

For any sheaves ${\cal S},\ {\cal T}$, the cup product 
\begin{displaymath}
  H^1({\cal S})\otimes H^1({\cal T})\to H^2({\cal S}\otimes {\cal T})
\end{displaymath}
is given in terms of Cech representatives $\omega$ and $\eta$ of elements of
$H^1({\cal S})$ and $H^1({\cal T})$ by
\begin{displaymath}
  \left(\omega\cup\eta\right)_{abc}=\omega_{ab}\otimes \eta_{bc},
\end{displaymath}
where restrictions to appropriate smaller open sets are understood and 
suppressed from the notation.  Note that $\omega\cup \eta$ is indeed a
cocyle if $\omega$ and $\eta$ are:
\begin{eqnarray*}
  \delta\left(\omega\cup\eta\right)_{abcd}&=&
\omega_{bc}\otimes\eta_{cd}-\omega_{ac}\otimes\eta_{cd}+
\omega_{ab}\otimes\eta_{bd}-\omega_{ab}\otimes\eta_{bc}\\
&=&\left(\omega_{bc}-\omega_{ac}\right)\otimes\eta_{cd}+\omega_{ab}\otimes
\left(\eta_{bd}-\eta_{bc}\right)\\
&=&-\omega_{ab}\otimes\eta_{cd}+\omega_{ab}\otimes\eta_{cd}=0,
\end{eqnarray*}
and similarly $\omega\cup\eta$ is a coboundary if either $\omega$ or  $\eta$
is a coboundary while the other is a cocycle.

To describe the map (\ref{wedge2in}) we note that any map
\begin{displaymath}
  {\cal O}(-2,-2)\to {\cal O}(-2,0)\oplus{\cal O}(-1,-1)^4\oplus
{\cal O}(0,-2)
\end{displaymath}
is given by multiplication with an element of
\begin{equation}
  \label{factor}
{\cal O}(0,2)\oplus{\cal O}(1,1)^4\oplus{\cal O}(2,0)
\end{equation}
which we now compute.  From (\ref{edual}) we compute that 
on the open set where
$\tilde{x}_1\ne0$, the sheaf ${\cal E}^\vee$ is the row space of
\begin{displaymath}
  \left(
  \begin{array}{cccc}
\tilde{x}_1 x_2 & -\tilde{x}_1 x_1& (\epsilon_2-\epsilon_1)x_1 x_2 & 0\\
0 & 0 & \tilde{x}_2 & -\tilde{x}_1
  \end{array}\right)
\end{displaymath}
with a similar expression when $\tilde{x}_2\ne0$.  The maximal minors of this
matrix are
\begin{displaymath}\left(
  \begin{array}{cccccc}
0 & x_2 \tilde{x}_1 \tilde{x}_2 & -\tilde{x}_1^2 x_2 & -x_1 \tilde{x}_1
\tilde{x}_2 & x_1\tilde{x}_1^2 & (\epsilon_2-\epsilon_1)x_1 x_2\tilde{x}_1    
  \end{array}\right).
\end{displaymath}
Note that this is a multiple of
\begin{equation}
  \label{thefactor}
\left(  \begin{array}{cccccc}
0 & x_2 \tilde{x}_2 & -\tilde{x}_1 x_2 & -x_1 
\tilde{x}_2 & x_1\tilde{x}_1 & (\epsilon_2-\epsilon_1)x_1 x_2
  \end{array}\right).
\end{equation}
We get precisely the same factor over the open set where $\tilde{x}_2\ne0$,
therefore this must be the desired element of (\ref{factor}).  We can
rewrite this as
\begin{equation}
\label{wedgefact}
x_2 \tilde{x}_2 e_1\wedge f_1 -\tilde{x}_1 x_2 e_1\wedge f_2 -x_1 
\tilde{x}_2 e_2\wedge f_1 +x_1\tilde{x}_1 e_2\wedge f_2 + 
(\epsilon_2-\epsilon_1)x_1 x_2  f_1\wedge f_2.
\end{equation}
This gives a computational simplification.  When we compute a cup and wedge
product of elements of $H^1({\cal E})$, then the resulting Cech representative
written as a representative of $H^2({\cal O}(-2,0)\oplus{\cal O}(-1,-1)^4
\oplus{\cal O}(0,-2))$ must be a multiple of (\ref{wedgefact}) on each
open set.  To find the multiple, hence the class in $H^2({\cal O}(-2,-2))$,
we need only compute the coefficient of one of the $e_i\wedge f_j$, say, 
$e_1\wedge f_1$.

Let's compute $\langle \tilde{Y}^2\rangle$.  Since there are no $e_i$
in the Cech representatives (\ref{cechyt}) for $\tilde{Y}_{ij,i'j'}$, 
we cannot obtain an $e_1\wedge
f_1$ term in the cup product.  Hence $\langle \tilde{Y}^2\rangle=0$.

For $\langle Y \tilde{Y}\rangle$ we compute using (\ref{cechyt}) and
(\ref{cechy}), omitting terms not involving
$e_1$ or $f_1$
\begin{eqnarray*}
\left( Y\cup\tilde{Y}\right)_{11,12,21}&=&
\left(\epsilon_1\tilde{x}_1^{-1}f_1\right)\wedge\left(-\tilde{x}_1^{-1}f_1
\right)=0\\
\left( Y\cup\tilde{Y}\right)_{11,21,22}&=&\left(-x_1^{-1}e_1+
\left(\epsilon_1\tilde{x}_1^{-1}-\epsilon_2\tilde{x}_2^{-1}\right)f_1\right)
\wedge
\left(-\tilde{x}_1^{-1}f_1\right)=\frac1{x_1\tilde{x}_1}e_1\wedge f_1\\
\left( Y\cup\tilde{Y}\right)_{11,12,22}&&=\left(
\epsilon_1\tilde{x}_1^{-1}f_1\right)\wedge 0=0\\
\left( Y\cup\tilde{Y}\right)_{12,21,22}&=&\left(
-x_1^{-1}e_1-\epsilon_2\tilde{x_1}^{-1}f_1\right)\wedge\left(
-\tilde{x}_1^{-1}f_1\right)=\frac1{x_1\tilde{x}_1}e_1\wedge f_1.
\end{eqnarray*}
After dividing by the coefficient $x_2\tilde{x}_2$ of $e_1\wedge f_1$ from
(\ref{wedgefact}), we learn that the Cech representative $\phi$ of $Y\tilde{Y}$
as an element of ${\cal O}(-2,-2)$ is given by
\begin{equation}
\label{therep}
  \phi_{11,12,21}=0,\qquad\phi_{11,21,22}=
\left(x_1x_2\tilde{x}_1\tilde{x}_2\right)^{-1},
\qquad \phi_{11,12,22}=0,\qquad\phi_{12,21,22}=
\left(x_1x_2\tilde{x}_1\tilde{x}_2\right)^{-1}.
\end{equation}

We now make an explicit choice of isomorphism
$H^2({\cal O}(-2,-2))\simeq{\bf C}$.  First of all, it is not hard to see
that Cech 
representatives of $H^2({\cal O}(-2,-2))\simeq{\bf C}$
that do not have a term involving $(x_1x_2\tilde{x}_1 
\tilde{x}_2)^{-1}$ are coboundaries.  For example,
a section of the form $(x_1^2\tilde{x}_1\tilde{x}_2)^{-1}$ can be extended
to $U_{11}\cap U_{12}$ and then used to construct a coboundary.  So we
only need to look at coefficients of $1/(x_1 x_2 \tilde{x}_1
\tilde{x}_2)$.  Let $A_{ij,i'j',i''j''}$ be these coefficients.  Note that the
cocycle condition implies
\begin{equation}
\label{coc}
  A_{12,21,22}-A_{11,21,22}+A_{11,12,22}-A_{11,12,21}=0.
\end{equation}

Now take a section $\rho$ on $U_{11}\cap U_{22}$ and compute
\begin{displaymath}
  \delta\rho_{11,12,22}=  \delta\rho_{11,21,22}=\rho.
\end{displaymath}
Similarly, if $\rho$ is a section on $U_{12}\cap U_{21}$ we compute
\begin{displaymath}
  \delta\rho_{11,12,21}=  \delta\rho_{12,21,22}=\rho.
\end{displaymath}
Since these generate all possible ways of getting terms $(x_1x_2
\tilde{x}_1^{-1}\tilde{x}_2)^{-1}$ in coboundaries we conclude that the
coboundaries satisfy
\begin{equation}
\label{cb2}
  A_{11,12,22}=  A_{11,21,22},\qquad
  A_{11,12,21}=  A_{12,21,22}.
\end{equation}

We therefore can define
\begin{equation}
  \label{tracemap}
{\rm tr}:H^2({\cal O}(-2,-2))\to {\bf C},\qquad {\rm tr}(\omega)=
A^\omega_{11,12,22}-  A^\omega_{11,21,22},
\end{equation}
where $A^\omega$ is the coefficient of $(x_1x_2
\tilde{x}_1^{-1}\tilde{x}_2)^{-1}$ in the Cech cocycle $\omega$.  Note
that this is the unique (up to a multiple) linear functional on the $A$'s
subject to the cocycle condition (\ref{coc}) which vanishes on coboundaries.

Finally, applying (\ref{tracemap}) to (\ref{therep}) we get
${\rm tr}(\phi)=1$.  Thus $\langle Y\tilde{Y}\rangle =1$.

We similarly compute the cup product of $Y$ with itself using (\ref{cechy})
and omitting terms not involving $e_1\wedge f_1$
\begin{eqnarray*}
  \left(Y\cup Y\right)_{11,12,21}&=&
\frac{\epsilon_1}{x_1\tilde{x}_1}e_1\wedge f_1\\
  \left(Y\cup Y\right)_{11,21,22}&=&
-\frac{\epsilon_2}{x_1\tilde{x}_1}e_1\wedge f_1\\
  \left(Y\cup Y\right)_{11,12,22}&=&
\frac{\epsilon_1}{x_1\tilde{x}_1}e_1\wedge f_1\\
  \left(Y\cup Y\right)_{12,21,22}&=&
-\frac{\epsilon_2}{x_1\tilde{x}_1}e_1\wedge f_1.
\end{eqnarray*}
Thus the Cech representative $\rho$ in $H^2({\cal O}(-2,-2)$ satisfies
\begin{displaymath}
  A^\rho_{11,12,21}=\epsilon_1,\ A^\rho_{11,21,22}=-\epsilon_2,\
  A^\rho_{11,12,22}=\epsilon_1,\ A^\rho_{12,21,22}=-\epsilon_2,
\end{displaymath}
leading to ${\rm tr}(\rho)=-(\epsilon_1+\epsilon_2)$.  Hence
$\langle Y^2\rangle=-(\epsilon_1+\epsilon_2)$.

Summarizing the classical computation, we have computed
\begin{displaymath}
  \langle \tilde{Y}^2\rangle = 0, \qquad
  \langle Y \tilde{Y}\rangle = 1, \qquad
  \langle {Y}^2\rangle = -(\epsilon_1+\epsilon_2).
\end{displaymath}
We now make the substitution
\begin{equation}
  \label{newsub}
X = Y+\epsilon_1\tilde{Y}, \qquad
\tilde{X} = \tilde{Y},
\end{equation}
which extends (\ref{substitution}) which held on the $(2,2)$ locus.
With these new variables, we once again get the usual classical 
correlators (\ref{classcorr}).

We now compute in the degree $(1,0)$ instanton sector.
As we have seen, there are several parts to giving a mathematical
formulation of the correlation functions.  First, we need to give a map
\[
\psi:H^p(X,\Lambda^q{\cal E}^{\vee})\to 
H^p({\cal M},\Lambda^q{\cal F}^{\vee}).
\]
To compute a correlation function $\langle \phi_1,\ldots,\phi_n\rangle$,
we start by multiplying the $\psi(\phi_i)$ (via cup and wedge products)
to get an element of
$H^4({\cal M},\Lambda^4{\cal F}^{\vee})$, so we next need to be able to
compute cup and wedge products.  Finally, we need to evaluate this class
numerically.  Since $\Lambda^4{\cal F}^{\vee}\simeq K_{{\cal M}}$, we have
$H^4({\cal M},\Lambda^4{\cal F}^{\vee})\simeq {\bf C}$.  
As in the classical calculation, we will use Cech cohomology and a trace map
\[
{\rm tr}:H^4({\cal M},\Lambda^4{\cal F}^{\vee})\to {\bf C}
\]

Then the correlation functions are
\begin{equation}
\label{correlation}
\langle \phi_1,\ldots,\phi_n\rangle_{1,0} =
{\rm tr}\left(\psi(\phi_1)\cup\cdots\cup\psi(\phi_n)\right).
\end{equation}
where the subscript emphasizes that we are looking only at instanton
contributions of a fixed degree.

As we have seen before, from the dual of (\ref{eq:e}) we have
$H^1(X,{\cal E}^{\vee})={\bf C}^2$ while from the dual of (\ref{eq:f})
we have $H^1({\cal M},{\cal F}^{\vee})={\bf C}^2$.
Explicitly the exact sequence
\begin{equation}
\label{eq:fd}
0\to {\cal F}^{\vee}\to {\cal O}(-1,0)^4\oplus{\cal O}(0,-1)^2
\to {\cal O}^2\to 0
\end{equation}
whose coboundary map
\[
H^0({\cal M},{\cal O}^2)\to H^1({\cal M},{\cal F}^{\vee})
\]
is an isomorphism by the vanishing of the cohomologies of
${\cal O}(-1,0)^4\oplus{\cal O}(0,-1)^2$.
Thus the identification of $H^1({\cal F}^{\vee})$ with ${\bf C}^2$
is canonical.  Similarly, the isomorphism of $H^1(X,{\cal E}^{\vee})$
with ${\bf C}^2$ is canonical.
Composing these canonical isomorphisms gives the desired isomorphism
\[
\psi:H^0(X,{\cal E}^{\vee})\simeq H^0({\cal M},{\cal F}^{\vee}).
\]
We need to explicitly evaluate (\ref{correlation}).  For this, we have
$\Lambda^4{\cal F}^{\vee}\simeq K_{{\cal M}}\simeq
{\cal O}(-4,-2)$.  So the fourth exterior
power of (\ref{eq:fd}) gives
\[
{\cal O}(-4,-2)\simeq\Lambda^4{\cal F}^{\vee}\hookrightarrow\Lambda^4\left(
{\cal O}(-1,0)^4\oplus{\cal O}(0,-1)^2\right)={\cal O}(-4,0)\oplus
{\cal O}(-3,-1)^8\oplus{\cal O}(-2,-2)^6.
\]
Note that this embedding is equivalent to giving a global section
of
\[
{\cal O}(0,2)\oplus{\cal O}(1,1)^8\oplus{\cal O}(2,0)^6.
\]
{}From (\ref{eq:fd}) and the explicit matrices in (\ref{eq:f}), we see
that over the open set where $\beta_1\ne 0$,
the rank 4 bundle ${\cal F}^{\vee}$ is spanned by the rows of the
matrix
\[
\left(
\begin{array}{cccccc}
\alpha_1&-\alpha_0&0&0&0&0\\
0&0&\alpha_1'&-\alpha_0'&0&0\\
0&0&0&0&\beta_1&-\beta_0\\
-\alpha_1'\beta_1&0&\alpha_0\beta_1&0&0&(\epsilon_1-\epsilon_2)\alpha_0
\alpha_1'
\end{array}
\right)
\]
Over the open set where $\beta_0\ne0$ the last row is replaced by a
similar expression.
The maximal minors of this matrix are computed to be
\[
0,\alpha_0{\alpha_1'}^2\beta_1^2,-\alpha_0\alpha_0'{\alpha_1'}\beta_1^2,\dots
\]
when written as the coefficients of $e_0\wedge e_1\wedge e_2\wedge e_3, 
e_0\wedge e_1\wedge e_2\wedge f_0,\ldots$.
Note that this is just $\alpha_0\alpha_1'\beta_1$ times
\begin{equation}
\label{subbundle}
0,{\alpha_1'}\beta_1,-\alpha_0'\beta_1,\dots
\end{equation}
and we would get exactly the same thing for the $\beta_0\ne0$
calculation.  This is in fact the ($\epsilon_i$-dependent) global section of
${\cal O}(0,2)\oplus{\cal O}(1,1)^8\oplus{\cal O}(2,0)^6$ that determines
$\Lambda^4{\cal F}^{\vee}$.

Let $\psi(Y)$ and $\psi(\tilde{Y})$ be the natural generators of $H^1({\cal M},
{\cal F}^{\vee})$.  Let's start by computing
$\langle \tilde{Y}^4\rangle_{1,0}$.
For simplicity, let's put $\alpha_2=\alpha_0'$ and $\alpha_3=\alpha_1'$.
We have the cover of ${\cal M}$ given by open sets $U_{ij}$ defined by
$\alpha_i\ne0,\beta_j\ne0$ (here $i=0,1,2,3$ and $j=0,1$.).  We have
$\psi(\tilde{Y})$ as the coboundary of $(0,1)\in H^0({\cal O}^2)$ in the exact
sequence (\ref{eq:fd}).  On any of the open sets $U_{ij}$ we lift
$(0,1)$ to the section $\beta_j^{-1}f_j$ of ${\cal O}(-1,0)^4
\oplus{\cal O}(0,-1)^2$, where $f_0,f_1$ are the natural generators
of ${\cal O}(0,-1)^2$.
Thus $\psi(\tilde{Y})\in H^1({\cal F}^{\vee})$, represented
as a Cech cocycle for this
cover, is given by
\[
(\psi(\tilde{Y}))_{ij;i'j'}=\beta_{j'}^{-1}f_{j'}-\beta_j^{-1}f_j.
\]
Since this representation of $\psi(\tilde{Y})$ lives entirely in the
rank 2 subbundle spanned by ${\cal O}(0,-1)^2$, the fourth exterior
powers in the cup product all vanish, and the Cech cocycle
representing
$\psi(\tilde{Y})\cup\psi(\tilde{Y})\cup\psi(\tilde{Y})\cup\psi(\tilde{Y})$ is
zero.  Hence

\begin{equation}
\label{y40}
\langle \tilde{Y}^4\rangle_{1,0}=0.
\end{equation}

Since the third exterior powers of the representatives for $\psi(\tilde{Y})$
similarly vanish, we immediately get

\begin{equation}
\label{y310}
\langle Y\tilde{Y}^3\rangle_{1,0}=0.
\end{equation}
Actually the second exterior powers vanish as well, since the representatives
live in the line subbundle of ${\cal O}(0,-1)^2$ which is the kernel of
the mapping
\[
{\cal O}(0,-1)^2\to {\cal O}
\]
with matrix $(\beta_0,\beta_1)$.  Thus

\begin{equation}
\label{y220}
\langle Y^2\tilde{Y}^2\rangle_{1,0}=0.
\end{equation}

Let us next compute $\psi(Y)\in H^1({\cal F}^\vee)$.  On $U_{ij}$ we
can lift $(1,0)$ to $\alpha_i^{-1}e_i-\epsilon_{c(i)}\beta_j^{-1}f_j$
on $U_{ij}$, where $e_0,e_1,e_2,e_3$ are local generators of ${\cal
O}(-1,0)^4$ and $c(i)=(1,1,2,2)$ for $i=0,1,2,3$ respectively.  This
gives
\begin{equation}
\label{cocycle}
(\psi(Y))_{ij;i'j'}=\alpha_{i'}^{-1}e_{i'}-\alpha_i^{-1}e_i
+\epsilon_{c(i)}\beta_j^{-1}f_j-\epsilon_{c(i')}\beta_{j'}^{-1}f_{j'}
\end{equation}
It remains to compute $\langle Y^3\tilde{Y}\rangle_{1,0}$
and $\langle Y^4\rangle_{1,0}$.
The straightforward way to
do this is to multiply the Cech cocycles found above and then use any
explicit form of the trace map.

The drawback is that there are many open sets needed to fully describe
Cech 4-cocycles, so the actual computation will be somewhat tedious.
Instead, we conclude by using some tricks adapted to this situation.

Since the coefficient of $e_0\wedge e_1\wedge e_2\wedge f_0$ (the
second entry in (\ref{subbundle}) is nonzero), we merely need to
extract the coefficient of $e_0\wedge e_1\wedge e_2\wedge f_0$ for the purpose
of computing.

In computing $\langle Y^3\tilde{Y}\rangle_{1,0}$, the only way to
get an $e_0\wedge e_1\wedge e_2\wedge f_0$ term is to use the $e_i$
from $\psi(Y)$ and the $f_0$ from $\psi(\tilde{Y})$.  Note that there is
no $\epsilon$ dependence in any of these terms.  Therefore $\langle
Y^3\tilde{Y}\rangle_{1,0}$ is independent of $\epsilon$.  Changing
basis to $X,\tilde{X}$ via (\ref{newsub}) and using the vanishings
(\ref{y40}), (\ref{y310}), and (\ref{y220}) we conclude that 
$\langle X^3\tilde{X}\rangle_{1,0}$ is independent of $\epsilon$.
By
putting $\epsilon_1=\epsilon_2$, it follows from the usual quantum cohomology
result that $\langle X^3\tilde{X}\rangle_{1,0}=1$.  Note also that
each Cech term in the cup product is a scalar multiple of
$(\alpha_0\alpha_1\alpha_2\alpha_3\beta_0\beta_1)^{-1}$ times
(\ref{subbundle}) so that the corresponding Cech cocycle in ${\cal
O}(-4,-2)$ is represented by
$(\alpha_0\alpha_1\alpha_2\alpha_3\beta_0\beta_1)^{-1}$, precisely of
the form needed to represent the nonzero class in $H^4({\cal M},
{\cal O}(-4,-2))$.

Turning to $\langle Y^4\rangle_{1,0}$, a similar argument
shows that the result is linear in the $\epsilon_i$, entering via the
term $\epsilon_{c(i)}\beta_0^{-1}f_0$ in (\ref{cocycle}).  But this
can be evaluated up to multiple by a simple trick.  
Note that via the change of basis (\ref{newsub}) and vanishings
(\ref{y40}), (\ref{y310}), (\ref{y220}) we infer that
$\langle X^4\rangle_{1,0}$ is linear in $\epsilon$.  Now recall that on
the $(2,2)$ locus, the correlation function $\langle X^4\rangle_{1,0}$
vanishes.  
Since the $(2,2)$ locus is described by the equation
$\epsilon_1-\epsilon_2=0$, we conclude that the correlation function
must simply be a multiple of $\epsilon_1-\epsilon_2$, i.e.
\[
\langle X^4\rangle_{1,0}\propto(\epsilon_1
-\epsilon_2).
\]

Summarizing, we have found
\begin{equation}
  \label{10results}
\langle \tilde{X}^4 \rangle_{1,0}=
\langle X\tilde{X}^3 \rangle_{1,0}=
\langle X^2\tilde{X}^2 \rangle_{1,0}=0,\ 
\langle X^3\tilde{X} \rangle_{1,0}=1,\
\langle {X}^4 \rangle_{1,0}\propto \epsilon_1-\epsilon_2.
\end{equation}
\smallskip
We can now repeat the computation for maps of degree $(0,1)$.  The linear sigma
model moduli space is ${\bf P}^1\times {\bf P}^3$.  We let
$\beta_0,\ \beta_1$ be the natural homogeneous coordinates on the
${\bf P}^1$ factor and let $\alpha_0,\ 
\alpha_1,\ \alpha_2,\ \alpha_3$ denote the homogeneous coordinates
on the ${\bf P}^3$ factor. Then the bundle ${\cal F}'$
on this moduli space is given by
\begin{equation}
\label{eq:fp}
0 \: \longrightarrow \:
{\cal O} \oplus {\cal O} \stackrel{ {\scriptsize
\left[ \begin{array}{cc}
       \beta_0 & \epsilon_1 \beta_0 \\
       \beta_1 & \epsilon_2 \beta_1 \\
       0 & \alpha_0\\
       0 & \alpha_1 \\
       0 & \alpha_2 \\
       0 & \alpha_3 \end{array} \right] }}{\longrightarrow} \:
{\cal O}(1,0)^2 \oplus {\cal O}(0,1)^4 
\: \longrightarrow \: {\cal F}' \: \longrightarrow \: 0.
\end{equation}
There is the dual exact sequence
\begin{displaymath}
0\to \left({\cal F}'\right)^\vee\to {\cal O}(-1,0)^2\oplus {\cal
O}(0,-1)^4\to {\cal O}^2\to 0.
\end{displaymath}
The section $(0,1)$ lifts to local sections $\alpha_j^{-1}e_j$ of
${\cal O}(-1,0)^2\oplus {\cal O}(0,-1)^4$ while $(1,0)$ lifts to
$\beta_i^{-1}f_i-\epsilon_{i+1}\alpha_j^{-1}e_j$ with $f_0,\ f_1,\
e_0,\ e_1,\ e_2,\ e_3$ basis vectors for ${\cal O}(-1,0)^2\oplus {\cal
O}(0,-1)^4$.  Over an open set, the rank 4 bundle $({\cal F}')^\vee$
is spanned by the rows of the matrix
\[
\left(
\begin{array}{cccccc}
0&0&\alpha_1&-\alpha_0&0\\
0&0&0&\alpha_2&-\alpha_1&0\\
0&0&0&0&\alpha_3&-\alpha_2\\
\alpha_0\beta_1&-\alpha_0\beta_0&(\epsilon_2-\epsilon_1)\beta_0\beta_1&0&0&0
\end{array}
\right)
\]
An argument analogous to the $(1,0)$ case says that we must have one
$f_i$ term and 3 $e_j$ terms in the wedge product arising from a
correlation function in order to get a nonzero answer.  Since there
are no $f_i$ in the lift of $(0,1)$, we immediately see that $\langle
\tilde{Y}^4 \rangle_{0,1}=0$.\footnote{We could instead have argued that
the representatives of the lifts live naturally in a rank 3
subbundle.}  

Similarly, from one $Y$ and three $\tilde{Y}'s$, the $f_i$ must come
from $Y$ and we see that there is no $\epsilon$ dependence.  We get
that $\langle X, \tilde{X}^3\rangle_{0,1}=1$ from the change of basis
(\ref{newsub}) the $(2,2)$ result.  

More generally, we see by the same reasoning that each four-point function
$< X^{i+1} \tilde{X}^{3-i}>_{0,1}$ is a homogeneous polynomial in the
$\epsilon_i$ of degree $i$, for $i=0,1,2,3$.  Furthermore,
each of these polynomials vanish on the (2,2) locus, hence is a multiple
of $\epsilon_1 - \epsilon_2$.

Summarizing, we have computed
\begin{equation}
  \label{01results}
\begin{array}{c}
\langle \tilde{X}^4\rangle_{0,1}=0,\ 
\langle X\tilde{X}^3\rangle_{0,1}=1,\ 
\langle X^2\tilde{X}^2\rangle_{0,1}\propto \epsilon_1-\epsilon_2,\\
\langle X^3\tilde{X}\rangle_{0,1}=(\epsilon_1-\epsilon_2)f_1(\epsilon),\ 
\langle X^4\rangle_{0,1}=(\epsilon_1-\epsilon_2)f_2(\epsilon),
\end{array}
\end{equation}
where $f_i(\epsilon)$ is a homogeneous polynomial of degree $i$ in
$\epsilon_1,\epsilon_2$.

Combining (\ref{10results}) with (\ref{01results}), we find
agreement with (\ref{xyyy}), (\ref{yyyy}), (\ref{xxyy}), (\ref{xxxy}),
and (\ref{xxxx}).

\section{Conclusions}

In this paper we have described the computation of
generalizations of the $\overline{ {\bf 27} }^3$ coupling in
perturbative heterotic string compactifications. 
These calculations amount to a heterotic version of curve-counting,
generalizing the standard A model calculations.

We spent the first third of this paper describing formally
how one can calculate these correlation functions.
We saw how old ideas from A model calculations generalize in the
heterotic context -- for example, the obstruction sheaf story
seems to generalize in an interesting way.

Our formal methods in the first third of this paper suffered
from needing not only a compactification of the moduli space
of worldsheet instantons, but also extensions of induced sheaves
over that compactification divisor, in a fashion that preserves
certain necessary properties of the Chern classes.
The second third of this paper was devoted to solving this
problem using linear sigma
models, which not only compactify moduli spaces, but as we
described, also provide
the needed sheaf extensions.

In the final part of this paper, we applied this technology to
check some predictions of \cite{abs} for heterotic curve counting.

There are several open problems that need to be solved:
\begin{enumerate}
\item We have described how to map 
\begin{displaymath}
H^p(X, \Lambda^q {\cal E}^{\vee} ) \: \mapsto \:
H^p({\cal M}, \Lambda^q {\cal F}^{\vee} )
\end{displaymath}
on open subsets of the moduli space, and in special cases involving
linear sigma models, have used calculational tricks to extend the
map over the compactification of the moduli space.
However, a more general prescription for extending the map 
over the compactification
is lacking.
\item In section~\ref{excesszeromodes}, we described a proposal for
generalizing obstruction sheaf constructions.  From physics, we conjecture
that
the Atiyah class of ${\cal E}$ determines an element of
\begin{displaymath}
H^1\left( {\cal M}, {\cal F}^{\vee} \otimes {\cal F}_1 \otimes
{\cal G}_1^{\vee} \right)
\end{displaymath}
with the property that when ${\cal E} = TX$, the element of the sheaf
cohomology group above is the Atiyah class of the obstruction sheaf.
With those assumptions, an easy Grothendieck-Riemann-Roch argument
showed how the resulting product of sheaf cohomology groups generated
a top form which can be integrated over the moduli space,
and we also checked that we reproduce the usual obstruction sheaf story
on the $(2,2)$ locus.
However, although the underlying physics seems clear,
these mathematical conjectures need to be checked.
\end{enumerate}

There are other extensions of this work that would be interesting
to pursue.  For example, it would be interesting to understand how
these correlation functions change when the K\"ahler class passes
through a stability subcone wall \cite{kcsub}, which would be the
heterotic analogue of a flop.

It would also be interesting to better understand the effect
of the ratio of operator determinants that we outlined earlier.
For the calculations in this paper, at genus zero, that ratio is just
a constant, which we have ignored.  However, at higher genus,
it is a nontrivial function of both the moduli of the Riemann surface
and of the bundle.

\section{Acknowledgements}

We would like to thank A.~Adams, J.~Bryan, R.~Plesser, and S.~Sethi for useful
conversations.  The work of SK has been
partially supported by NSF grant DMS 02-96154 and NSA grant
MDA904-03-1-0050.  The work of ES has been partially supported by NSF
grant DMS 02-96154.

\end{document}